\documentclass[12pt,a4paper]{article}
\pdfoutput=1
\usepackage{jinstpub}

\usepackage{adjustbox}
\usepackage{isotope} % \isotope command

\usepackage{scrextend}

\usepackage{amsmath}
\usepackage{amssymb}
\usepackage[nowatermark]{fixmetodonotes}
\usepackage{comment}
\usepackage{siunitx}
\usepackage{tikz}
\usetikzlibrary{patterns}
\DeclareSIUnit[number-unit-product = ]\percent{\char`\%}
\DeclareSIUnit\foot{ft}
\DeclareSIUnit\inch{inch}
\DeclareSIUnit\pot{POT}
\DeclareSIUnit\year{years}
\DeclareSIUnit\ADC{ADC\ counts}
\DeclareSIUnit\spill{spill}
\DeclareSIUnit\ton{metric~tons}

\usepackage{booktabs}

\usepackage{cleveref}

\crefname{figure}{figure}{figures}
\Crefname{figure}{Figure}{Figures}

\crefname{table}{table}{tables}
\Crefname{table}{Table}{Tables}

\crefname{section}{section}{sections}
\Crefname{section}{Section}{Sections}

\usepackage{makecell}
\usepackage{multirow}
\usepackage[inline]{enumitem}
\usepackage{pgfplots}
\usepackage{pgfplotstable}

\pgfplotsset{
  axis line style={black},
  tick style={thick, black},
  every axis label/.append style={black},
  every tick label/.append style={black}
}

\newcommand{\phasei}{Phase-I}
\newcommand{\phaseii}{Phase-II}
\newcommand{\modea}{Mode A}
\newcommand{\modeb}{Mode B}

\newcommand{\tankRateEquation}{$\mathcal{R}_\text{n}^\text{tank} =
0.053\, ^{+0.053}_{-0.025}\vcenter{\hbox{\footnotesize\text{ stat+syst}}}$}

\pgfplotsset{error bar legend/.style={%
    /pgfplots/legend image code/.prefix code={%
      \pgfkeysgetvalue{/pgfplots/error bars/error mark}{\pgfplotserrorbarsmark}%
      \draw[%
        /pgfplots/every error bar,
        mark=none%\pgfplotserrorbarsmark,
        /pgfplots/error bars/error mark options,
        sharp plot,
      ] plot coordinates {(0.3cm, -0.15cm) (0.3cm, 0.15cm)};%
    }
  }
}

\pgfplotsset{filter discard warning=false}

\usepackage{graphicx}
\usepackage{amssymb}
\usepackage{lineno}

\begin{document}

\definecolor{MyGreen}{RGB}{31,116,48}
\definecolor{MyViolet}{RGB}{131,16,148}

%\begin{frontmatter}


\title{Measurement of Beam-Correlated Background Neutrons from the Fermilab Booster Neutrino Beam in ANNIE \phasei}

\author[b]{A.~R.~Back,}
\author[d]{J.~F.~Beacom,}
\author[c,1]{T.~Boschi,\note{Present address: Queen Mary University of London; London, E14NS, UK}}
\author[b]{D.~Carber,}
\author[b,2]{E.~Catano-Mur,\note{Present address: College of William \& Mary; Williamsburg, VA 23187, USA}}
\author[f,3]{M.~Chen,\note{Present address: Lehigh University; Bethlehem PA 18015, USA}}
\author[i]{E.~Drakopoulou,}
\author[c]{F.~Di~Lodovico,}
\author[h]{A.~Elagin,}
\author[b]{J.~Eisch,}
\author[f]{V.~Fischer,}
\author[f,4]{S.~Gardiner,\note{Present address: Fermi National Accelerator Laboratory; Batavia, IL 60510, USA}}
\author[g]{J.~Griskevich,}
\author[b,5]{D.~Grzan,\note{Present address: University of California, Davis; Davis, CA 95817, USA}}
\author[a]{R.~Hatcher,}
\author[b]{F.~Krennrich,}
\author[f]{B.~Kimmelman,}
\author[a]{A.~Kreymer,}
\author[a,6]{W.~Lee,\note{Deceased}}
\author[g]{S.~Locke,}
\author[f]{M.~Long,}
\author[j]{M.~Malek,}
\author[a]{C.~McGivern,}
\author[f]{E.~Moore,}
\author[i]{M.~Needham,}
\author[j]{M.~O'Flaherty,}
\author[b,7]{J.~Podczerwinski,\note{Present address: University of Wisconsin; Madison, WI 53708, USA}}
\author[k]{B.~Richards,}
\author[g]{J.~Ritz,}
\author[b]{M.~C.~Sanchez,}
\author[g]{M.~Smy,}
\author[f]{R.~Svoboda,}
\author[b]{E.~Tiras,}
\author[g]{M.~Vagins,}
\author[f]{J.~Wang,}
\author[g,8]{P.~Weatherly,\note{Present address: Drexel University; Philadelphia PA 19104, USA}}
\author[b]{A.~Weinstein,}
\author[b,9]{M.~Wetstein,\note{Corresponding author: \href{mailto:wetstein@iastate.edu}{wetstein@iastate.edu}}}
\author[b]{and J.~Wu}

%\emailAdd{wetstein@iastate.edu}

\affiliation[a]{Fermi National Accelerator Laboratory;
  \textit{Batavia, IL 60510, USA}}
\affiliation[b]{Iowa State University; \textit{Ames, IA 50011, USA}}
\affiliation[c]{Kings College London; \textit{London WC2R2LS, UK}}
\affiliation[d]{Ohio State University; \textit{Columbus, OH 43210, USA}}
\affiliation[e]{Queen Mary University of London; \textit{London E14NS, UK}}
\affiliation[f]{University of California, Davis; \textit{Davis, CA 95817, USA}}
\affiliation[g]{University of California, Irvine; \textit{Irvine, CA 92697,
USA}}
\affiliation[h]{University of Chicago, Enrico Fermi Institute;
  \textit{Chicago, IL 60637, USA}}
\affiliation[i]{University of Edinburgh; \textit{Edinburgh EH9 3FD, UK}}
\affiliation[j]{University of Sheffield; \textit{Sheffield S10 2TN, UK}}
\affiliation[k]{University of Warwick; \textit{Coventry CV47AL, UK}}

\collaboration{ANNIE Collaboration}

\abstract{
The Accelerator Neutrino Neutron Interaction Experiment (ANNIE) aims to make a
unique measurement of neutron yield from neutrino-nucleus interactions and to
perform R\&D for the next generation of water-based neutrino detectors. In this
paper, we characterize beam-induced neutron backgrounds in the experimental hall at Fermi National Accelerator Laboratory. It is shown that the background levels are sufficiently low to allow the next stage of the experiment to proceed. These measurements are relevant to other Booster Neutrino Beam (BNB) \cite{BNBTDR} experiments located adjacent to ANNIE Hall, where dirt neutrons and sky-shine could present similar backgrounds.
}

\maketitle


%\end{frontmatter}

%\linenumbers

\section{Introduction}
\label{sec:intro}

The Accelerator Neutrino Neutron Interaction Experiment (ANNIE)
\cite{ANNIEProposal} aims to make the first detailed measurement of the number
of neutrons produced by muon neutrinos interacting with nuclei. Measurements of the
final-state neutron multiplicity are key to improving our understanding of
neutrino-nucleus interactions. This in turn improves our understanding of
systematic uncertainties in neutrino oscillation experiments, where the energy
carried by difficult-to-detect final-state neutrons can degrade the resolution
of the reconstructed neutrino energy. Identifying and counting final-state
neutrons also provides a new and critical handle on signal-background separation
in future proton decay and neutrino experiments~\cite{FSneutrons}.

\begin{figure}[t]
        \begin{center}
           \begin{tabular}{c}
                        \includegraphics[width=0.7\linewidth]{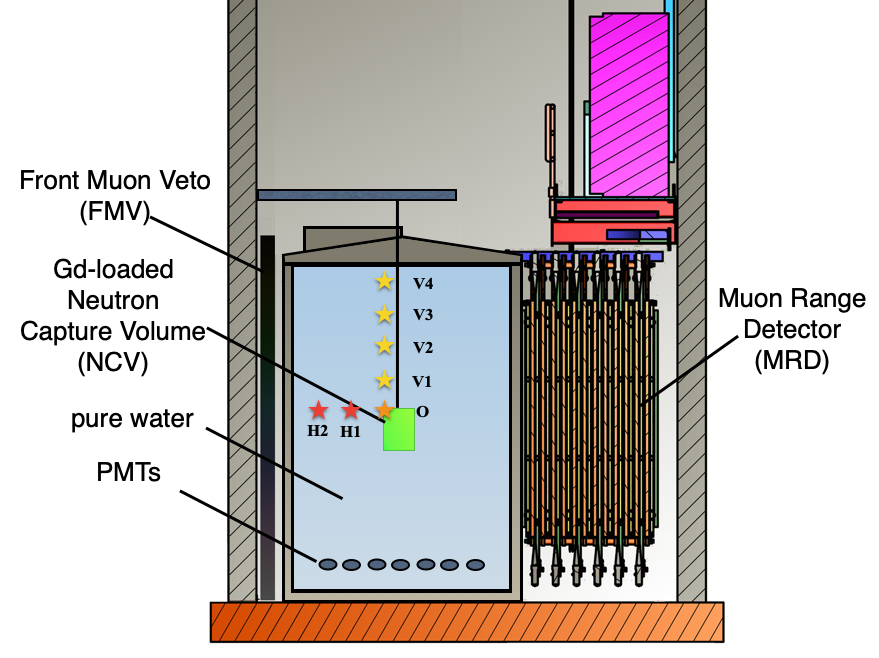}\\
                        \includegraphics[width=0.7\linewidth]{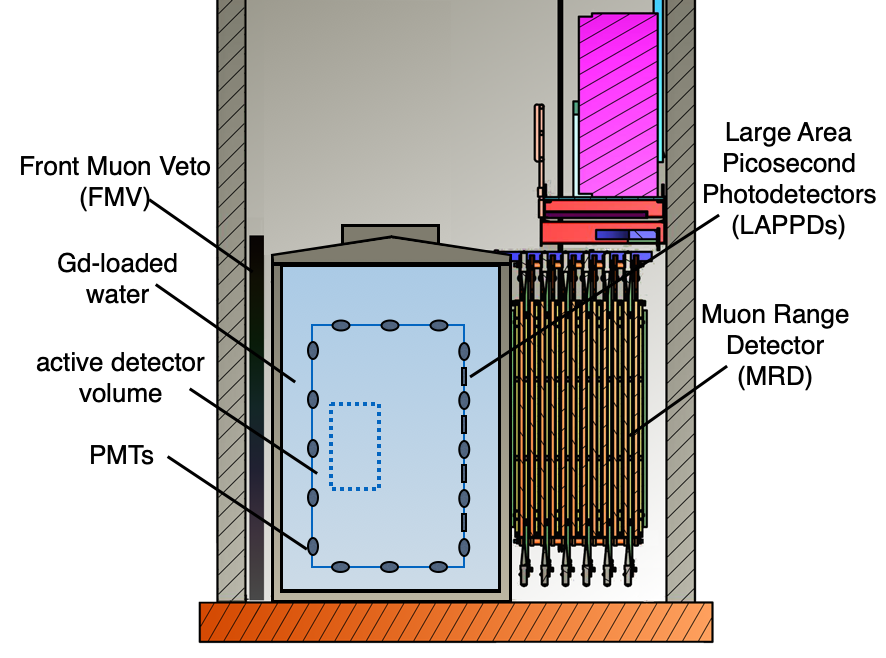}
                     \end{tabular}
        \end{center}
        \caption{ TOP:~A concept drawing of the \phasei\
ANNIE detector system, showing the positions of the upper left corner of the
neutron capture volume (NCV) described in \cref{sec:exp_design}. 
BOTTOM:~A concept drawing of the complete \phaseii\ detector. The
solid blue line indicates the optically isolated active volume of the detector
and the dotted blue line indicates the fiducial volume optimized for the
\phaseii\ physics measurement. }
\label{fig:anniedetector}
\end{figure}

\begin{figure}
\begin{center}
\includegraphics{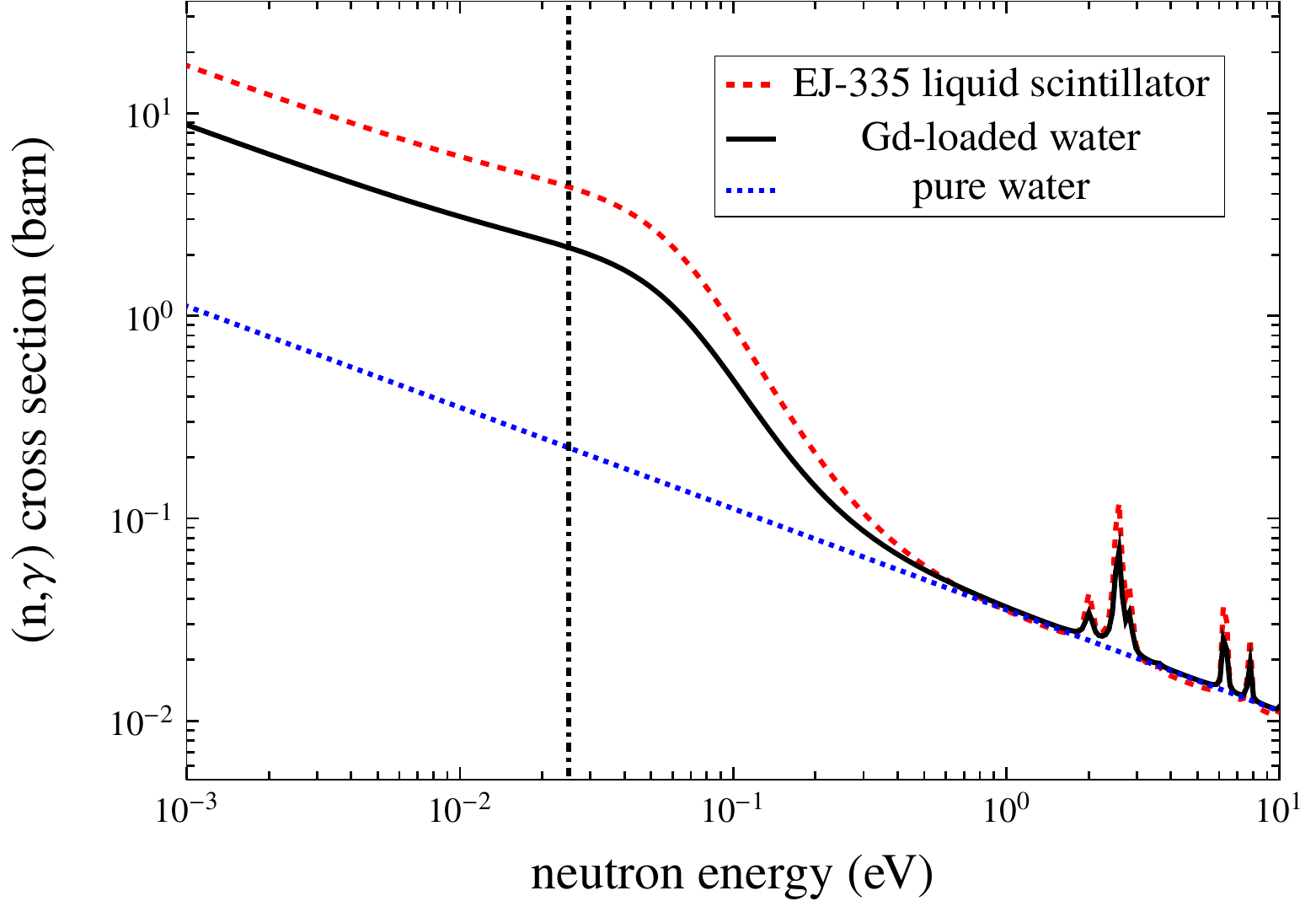}
\caption{The neutron capture cross section as a function of energy for pure
  water, EJ-335 Gd-loaded liquid scintillator (\SI{0.25}{\percent} w/w), and the Gd-loaded water (\SI{0.1}{\percent} w/w) to be used in ANNIE \phaseii. The results shown in the plot, which
  are weighted by nuclide fraction, were computed using cross sections for pure
  nuclides taken from ENDF/B-VIII.0~\cite{BROWN20181}. The
  dash-dotted vertical line at \SI{0.025}{\eV} indicates a typical
  kinetic energy for a thermal neutron.}
\label{fig:neutron_xsec}
\end{center}
\end{figure}

The lower panel of \cref{fig:anniedetector} shows the detector configuration (referred to as \phaseii) used to perform the final-state neutron multiplicity measurement. The main target consists of an upright cylindrical steel tank filled with \SI{26}{\ton} of gadolinium-loaded (Gd-loaded) ultra-pure deionized water, instrumented with photodetectors and partially enclosed by a muon detection system.
A muon produced by a neutrino interaction in the fiducial volume is reconstructed using the tank photodetectors and muon detection system. Neutrons produced by the neutrino interaction scatter and lose energy through thermalization, allowing them to capture on either H or Gd in the active volume. The fall-off in the neutron capture cross-section with energy for both pure and Gd-loaded water is shown in \cref{fig:neutron_xsec}. Gd-loading dramatically enhances the cross-section relative to pure water for energies near and below the thermal neutron energy of \SI{0.025}{\eV}.
The capture produces a delayed signal in the form of a de-excitation $\gamma$-ray cascade, with properties determined by the capturing nuclide. In particular, captures on Gd produce a more easily-detected $\sim$\SI{8}{\MeV} $\gamma$-ray cascade compared with the \SI{2.2}{\MeV} cascade from H-capture. At concentrations of \SI{0.1}{\percent}  Gd by mass, the enhanced cross-section has the added benefit of shortening the time constant for neutron capture from $\sim$\SI{200}{\micro\second} to $\sim$\SI{30}{\micro\second}. These combined effects make Gd-loading essential to the final-state neutron multiplicity measurement.

Because final-state neutrons in ANNIE can travel over a meter before thermalizing and capturing in the $\sim$\SI{14}{\meter\cubed} ANNIE active volume, neutrino-induced neutrons and background neutrons are spatially indistinguishable.
As a consequence, any neutrons entering the tank that are unrelated to the
neutrino interaction constitute a potential background for the ANNIE physics program.
Restricting the analysis to a narrow time window around the arrival of the neutrino beam spill suppresses a large fraction of the constant-in-time (CIT) background activity arising from processes unrelated to the beam. The residual CIT background can be characterized by taking off-beam triggers.
All remaining backgrounds are correlated in time with the beam.

There are two dominant types of beam-correlated neutron backgrounds, both of which are delayed relative to the prompt component from
beam neutrino interactions. The first type, referred to as \textit{sky-shine}, consists
of secondary neutrons produced in the beam dump that leak into the atmosphere
and enter the detector after undergoing multiple
scattering~\cite{LeeSkyshine,msu-skyshine}. Preliminary results from the SciBooNE experiment, which previously occupied the ANNIE experimental hall, show an excess of presumed sky-shine events after the beam spill with a clear dependence on detector depth~\cite{Takeiskyshine}. The
dependence of the event count with respect to depth suggests that using a
fiducial volume away from the top of the detector would significantly reduce the
sky-shine background. The second type of background, known as \textit{dirt neutrons},
consists of neutrons that arise from beam neutrino interactions occurring in the
dirt and rock upstream of the experimental hall. 
The optically isolated buffer region of water upstream of the ANNIE \phaseii\/ active volume should reduce the dirt neutron flux. 

In this paper, we report a first measurement of beam-correlated background neutrons in the ANNIE tank as a function of position. This analysis uses data taken in a special configuration (ANNIE \phasei) of the detector with a pure water target. This configuration is pictured schematically in the upper panel of \cref{fig:anniedetector}. We measure a rapid fall-off of the neutron background with depth and demonstrate that neutron backgrounds in the detector volume are limited. 
We also use the measured fall-off as a function of distance from the surface of the water and the tank walls in \phasei\ to verify that that the proposed buffer region surrounding the optically isolated volume for the main ANNIE neutrino interaction physics program, known as \phaseii\ (lower panel of \cref{fig:anniedetector}), provides adequate shielding from background neutrons.  As the size of this buffer region could only be increased by reducing the size of the optically isolated volume, this indicates that we can achieve the necessary background levels while accommodating a neutrino vertex fiducial volume large enough ($\sim$ \SI{2.5}{\cubic\metre}) to contain the neutrons from neutrino interactions.
These results establish the feasibility of the \phaseii\ physics program.

The measurements presented in this paper are relevant to other Booster Neutrino Beam (BNB) \cite{BNBTDR} experiments such as the Short Baseline Near Detector (SBND), located adjacent
to ANNIE Hall, where dirt neutrons and sky-shine could present similar
backgrounds. The techniques described in this paper will also be applicable to
any future water-based near detectors, especially those with Gd-loading
or water-based liquid scintillators.

\section{Experimental design of the  neutron background measurement}
\label{sec:exp_design}

The ANNIE detector is installed in the BNB at Fermilab
at the former location of the SciBooNE \cite{Kurimoto2009} detector. The BNB runs at an average rate of \SI{5}{\hertz}. 
Protons are delivered in 84 bunches over a \SI{1.6}{\micro\second} spill time to a target and horn combination
\SI{100}{\meter} upstream of ANNIE Hall. 
The nominal number of protons-on-target (\si{\pot}) per spill is \num{5e12} \si{\pot}. 
The beam is estimated to produce \SI{93}{\percent} pure $\nu_{\mu}$, with an energy spectrum peaking at around \SI{700}{\MeV}~\cite{BNBflux}\/.

The ANNIE \phasei\ neutrino target and optical instrumentation (shown in \cref{fig:anniedetector}) were contained in a steel tank roughly \SI{3}{\meter} in diameter by \SI{4}{\meter} in height. The interior of the tank was covered with a white reflective PVC liner in order to maximize light collection and was filled with \SI{26}{\ton} of ultra-pure deionized water. An array of 58 upward-facing \SI[number-unit-product=\text{-}]{8}{\inch} Hamamatsu R5912 photomultiplier tubes (PMTs) was installed inside the water volume at the base of the tank. All of these PMTs were mounted on an octagonal, stainless-steel inner structure that, along with the attached tank lid, could be lifted out and replaced without moving the tank itself. A set of six plastic scintillator paddles, each with an attached light guide and a \SI[number-unit-product=\text{-}]{2}{\inch} PMT, was mounted in a metal frame placed on top of the tank lid. These paddles were used to generate triggers on directionally-selected cosmic muons.

\subsection{Neutron Capture Volume (NCV)}

The position dependence of beam-correlated neutron backgrounds is
characterized using a movable Neutron Capture Volume (NCV) deployed within
the tank. The NCV is a \SI{50}{\centi\meter} $\times$ \SI{50}{\centi\meter} acrylic cylinder filled
with EJ-335, a Gd-loaded (0.25\% w/w) liquid scintillator manufactured by
Eljen Technology~\cite{ej-335}. Thermal neutron capture on Gd produces
a $\gamma$-ray cascade with a total energy of around \SI{8}{\MeV}, which is
detectable as a bright flash of light in the scintillator. The radiation length in
the NCV is roughly \SI{50}{\centi\meter}, and thus neutron captures are often not
fully contained, limiting the detection efficiency of the volume to around
\SI{10}{\percent} (see \cref{sec:ncv_efficiency}). The NCV is moved within the
water volume using a sliding winch. A slot on the hatch of the tank lid permits
translation of the NCV in the beam direction. All of the data used in this paper
were taken in a mode where the NCV was wrapped in successive layers of reflective white plastic to maximize total internal reflection and black plastic to optically isolate it from the rest of the tank. Two \SI[number-unit-product=\text{-}]{3}{\inch} PMTs were installed on top of the NCV in
order to tag energy depositions in the liquid scintillator.

\subsection{Upstream and downstream veto and muon selection}

A front muon veto (FMV) consisting of two layers of overlapping scintillator paddles (originally used by the CDF experiment~\cite{cdfpaddles}) sits between the tank and the beam. The FMV is used to reject charged particles produced in the dirt and rock upstream of the detector. A  muon range detector (MRD) consisting of 11 alternating layers of iron absorber and vertical and horizontal plastic scintillator paddles (previously used by SciBooNE~\cite{Hiraide:2008eu}) sits downstream from the neutrino target. For \phasei, only two layers were instrumented, which was sufficient to tag outgoing muons. From simulation studies using the GENIE generator~\cite{genie1}, it is estimated there are approximately \num{26000} charged-current muon neutrino interactions within the $\sim$\SI{2.5}{\cubic\metre} fiducial volume per year, of which roughly \num{5000} produce muons that enter and range-out in the MRD.

\subsection{Electronics, trigger, and data acquisition system}
\label{sec:daq}

\begin{figure}[t]
        \begin{center}
           \begin{tabular}{c}
                        \includegraphics[width=0.95
\linewidth]{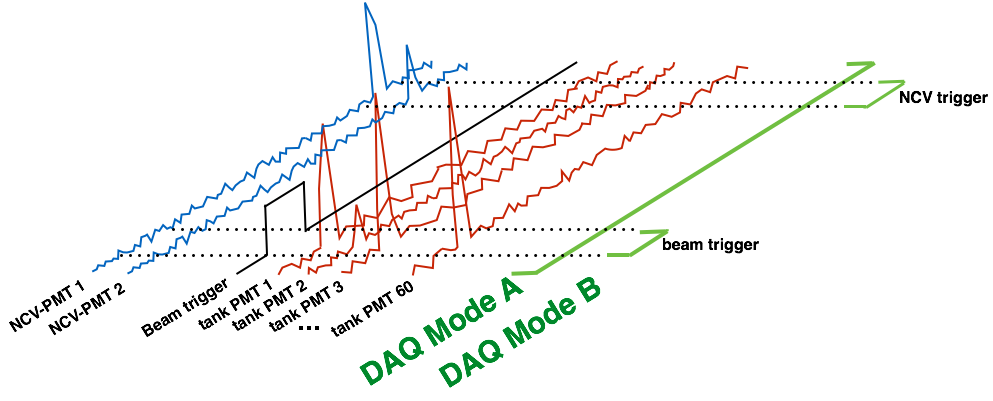}\\
                        \end{tabular}
        \end{center}
        \caption{A cartoon representation showing how a digitized ANNIE event is
recorded in the two different DAQ modes.
        }
                \label{fig:daqmodes}
\end{figure}

The detector electronics readout system consists of three subsystems. A Central Trigger Card (CTC) provides synchronization, time-stamping and event tagging. A VME-based system, originally designed for the K0TO experiment~\cite{K0TODAQ}, digitizes the full waveforms from all of the water PMTs and NCV PMTs at 500 MSamples/sec into a deep buffer capable of recording up to \SI{80}{\micro\second}. The VME system is also responsible for generating the triggers from the NCV and water PMTs. Finally, a CAMAC-based TDC system time stamps and records pulses above threshold from the FMV and the MRD.

These systems are integrated using the data acquisition (DAQ) framework ANNIEDAQ
\cite{benjamin_richards_2018_1489338}, a modular and scalable DAQ framework
based on ToolDAQ \cite{benajmin_richards_2018_1482767}. The software runs in a
distributed way on multiple servers and the VME computer cards. It is
responsible for run and high voltage control, slow-control monitoring, maintaining the run status database, trigger mode configuration,  and managing data from each of the asynchronously-running detector subsystems (FMV, water and NCV PMTs, and the MRD). The electronics and DAQ software are highly scalable and configurable and will serve as the baseline for the ANNIE Phase-II detector.

All trigger signals are managed by the configurable CTC, a CAEN V1495
general purpose VME board with ECL and LVDS inputs and a customizable FPGA. The
CTC receives a signal from the BNB facility that provides advance
notification of a beam spill arriving (beam trigger) and forwards appropriately
time-delayed copies to the VME and CAMAC systems. The CTCalso passes through and timestamps a Global Positioning System (GPS) 1PPS signal, used for synchronization.

Data taking occurred in two different trigger modes. These two modes are depicted in \cref{fig:daqmodes}. \modea\ records the waveforms for all channels for an \SI{80}{\micro\second} time window, large enough to include all prompt activity and the majority of subsequent neutron captures. The start of the beam trigger is placed \SI{10}{\micro\second} into the \SI{80}{\micro\second} buffer, providing a high statistics sample of the pre-beam random background. This mode has the advantage of being insensitive to the detection threshold for neutron captures.
\modeb\ records all channels during a \SI{2}{\micro\second} window around the beam trigger. Additional \SI{2}{\micro\second} recordings of all channels are stored for any NCV triggers (defined as the OR of the two NCV PMT signals) within a \SI{100}{\micro\second} gate following the beam trigger. An onboard 64-bit counter, calibrated using synchronization pulses from the central trigger card and the VME CPU UTC time, permits offline correlation of all ADC records associated with a Mode B trigger. This mode reduced deadtime while requiring a more sophisticated understanding of the relationship between neutron captures and trigger thresholds. It also required the adoption of pre-scaled off-beam triggers.

\section{Data taking and selection}
\label{sec:data}

The beam data used in this paper were collected from February through August of 2017, representing a total exposure of \num{6.24e19} \si{\pot}. Data were collected with the NCV at seven different positions,
shown in the top panel of \cref{fig:anniedetector} and using a mix of the two DAQ modes described in \cref{sec:daq}.
The NCV positions
are numbered with respect to a reference position (position O) at the center of the tank. The H positions are numbered to increase with decreasing horizontal shielding. The V positions are numbered to increase with decreasing water overburden. An uncertainty of \SI{2}{\centi\meter} was estimated for all NCV position measurements.

All beam data at position V4 were taken in DAQ mode A. All data at positions V1--3 and H1--2 were taken in DAQ mode B. At position O, located in the center of the tank, total exposures of \SI{3.57e18}{\pot} (\num{875867} beam spills) and \SI{1.409e19}{\pot} (\num{3692460} beam spills) were obtained in DAQ modes A and B, respectively. The measured rates were consistent within one-sigma uncertainties between the two approaches. In addition to the data in \cref{tab:data}, six \isotope[252]{Cf} source calibration runs were performed with the NCV at position V4, providing a total of \num{206732} source triggers.

\Cref{tab:data} summarizes the number of beam spills, total exposure, and total number of cosmic muon triggers recorded at each position. 

\begin{table}
\centering
\begin{minipage}{\textwidth}
\renewcommand{\footnoterule}{\vspace{-5pt}}
\caption{Summary of the data collected during ANNIE \phasei.}
\label{tab:data}
\smallskip
\renewcommand\cellset{\renewcommand{\arraystretch}{0.8}}
\setlength\tabcolsep{4pt} % default value: 6pt
\begin{tabular}{cSSrS[table-format=1.4, round-mode=places, round-precision=2]
S[table-format=1.4, round-mode=places, round-precision=2]r}
\toprule
{ \multirow{3}{*}{\makecell[c]{NCV \\ position}} }
& { \multirow{3}{*}{\makecell[c]{Water \\ overburden\footnote{The
thickness of water above the NCV.} \\ (\si{\centi\meter})}} }
& { \multirow{3}{*}{\makecell[c]{Water \\ shielding\footnote{The thickness of
water between the beam-side wall of the tank and the NCV.} \\ ($\si{\centi\meter}$) }} }
& { \multirow{3}{*}{\makecell[c]{Beam \\ spills}} }
& { \multirow{3}{*}{{\makecell[c]{Total \\ exposure \\ (\SI{e18}{\pot})}}} }
& { \multirow{3}{*}{{\makecell[c]{Average \si{\pot} \\ per spill
\\ (\SI{e12}{\pot})}}} }
& { \multirow{3}{*}{\makecell[c]{Cosmic \\ triggers}} } \\
\\
\\
\addlinespace[-3\aboverulesep]
\midrule
O  & 138(2) & 104(2) & \num{4568327} & 17.664  & 3.86662 & \num{33437} \\ \hline
H1 & 138(2) &  58(2) & \num{650378} & 2.20311 & 3.38743 & \num{6575} \\
H2  & 138(2) &  10(2) & \num{4383135} & 13.2434 & 3.02144 & \num{40451} \\ \hline
V1 &  67(2) & 104(2) & \num{2023082} & 6.79032 & 3.35642 & \num{26610} \\
V2 &  36(2) & 104(2) & \num{3476203} & 10.9768 & 3.15771 & \num{59387} \\
V3 &  21(2) & 104(2) & \num{973057} & 3.63194 & 3.73251 & \num{11502} \\
V4 &   6(2) & 104(2) & \num{1779098} & 7.88172 & 4.43017 & \num{17217} \\
\bottomrule
\end{tabular}
\end{minipage}
\end{table}

\section{Signal processing}

The \phasei\ data were processed and analyzed using ToolAnalysis~\cite{ToolAnalysis}, an event reconstruction software package developed by ANNIE collaborators within the ToolDAQ framework~\cite{benajmin_richards_2018_1482767}. All of the information used in neutron candidate reconstruction and selection is derived from PMTs attached to the NCV and the bottom of the tank. In this section we describe the PMT signal processing.

\subsection{Pedestal estimation}

The pedestal ADC value for each PMT channel, denoted $x_0$, is estimated according to the ZE3RA algorithm~\cite{ZE3RA}. To apply this algorithm, 40 time slices, each 25 ADC samples long, are chosen from regions of data that are expected to be pulse-free. For \modea, the 40 time slices are contiguous and correspond to the first \SI{2}{\micro\second} of the full \SI[number-unit-product=\text{-}]{80}{\micro\second} readout window.
For \modeb, the time slices are obtained from the first \SI{50}{\nano\second} of 40 consecutive
\SI[number-unit-product=\text{-}]{2}{\micro\second} data records.

The mean and variance of the 25 ADC values in each time slice are calculated, and the sample variances of neighbouring slices are checked for statistical consistency using an F-test. Time slices that are identified as inconsistent, likely due to the presence of nuisance pulses or electrical transients, are flagged and removed. The pedestal $x_0$ is then taken to be the mean of all ADC values from the remaining slices. The standard deviation $\sigma_{x_0}$ of the ADC samples selected in this way is used in PMT pulse finding.

\subsection{Pulse finding}
\label{sec:PMT_pulses}

To ensure threshold consistency between \modea\ and \modeb\ data, pulse finding is performed prior to pedestal subtraction and calibration for each PMT channel. Starting with the first digital sample in a waveform, ADC values are sequentially checked until the appropriate pulse finding threshold is exceeded. This corresponds to 357 ADC counts for the NCV PMTs and 7 ADC counts above pedestal for the tank PMTs.\footnote{The 357 ADC count threshold was chosen to permit its implementation in hardware rather than software for \modeb\/.}
The first digital sample for which this occurs is defined as the beginning of a pulse. The
subsequent samples are checked until one of the following conditions is met:
\begin{enumerate*}[label=(\arabic*)] \item an ADC value is found that falls
below $x_0 + \sigma_{x_0}$, or \item the end of the record is reached.
\end{enumerate*}
The digital sample fulfilling the logical OR of these criteria is defined to be the end of the pulse.

\subsection{Waveform calibration}

After subtracting the pedestal, the ADC waveforms        are calibrated.
For the ANNIE ADC cards, the voltage $V_\text{PMT}$ corresponding to
a recorded ADC value $x_\text{ADC}$ is given by the relation: \begin{equation}
\label{eq:calibration_formula} V_\text{PMT} = \frac{ \SI{2.415}{\volt} }{
\SI[exponent-base=2]{e12}{\ADC} } \, (x_\text{ADC} - x_0). \end{equation}  Calibrated waveforms are obtained by applying the conversion formula \cref{eq:calibration_formula} to
each of the raw samples.

\subsection{Feature extraction}

The fully calibrated pulses are characterized using the following feature-extraction parameters: \begin{enumerate*}[label=(\arabic*)] \item {\normalfont start time:} the
starting sample index multiplied by the sampling period; \item {\normalfont peak time:} the time at which the maximum ADC value occurred within the pulse (If the maximum ADC value was reached more than once
during the pulse, then the earliest sample for which this occurred is used to
calculate the peak time); \item {\normalfont end time:} the time corresponding to the
first sample after the start of the pulse at which either the ADC signal fell
below $x_0 + \sigma_{x_0}$ or the data record ended; \item {\normalfont raw amplitude:}
the maximum ADC value recorded during the pulse; \item {\normalfont calibrated amplitude:}
the maximum voltage recorded during the pulse, calculated using
\cref{eq:calibration_formula} with $x_\text{ADC}$ set equal to the raw amplitude
of the pulse; and \item {\normalfont charge:} the time integral of the calibrated version
of the pulse divided by the ADC input impedance (\SI{50}{\ohm}).\end{enumerate*}

\section{Neutron candidate selection}
\label{sec:eventselection}

Neutrons are identified by a burst of scintillation light detected within the NCV over a time region of interest between \SIrange[range-phrase=~and~]{10}{70}{\micro\second} after the beam arrival. Events are selected that have no prompt neutrino interaction in the tank but that have a delayed signal consistent with the capture of neutrons entering from outside the detector volume. The compact size of the NCV provides localization of the neutrons. The quantity and spread of the light detected on the PMTs at the bottom of the water volume are used to veto cosmic muons. The count of neutrons detected in the NCV is divided by the beam exposure and NCV volume to determine the rate of background neutrons per unit volume per beam spill.

\begin{figure}
\begin{center}
\includegraphics{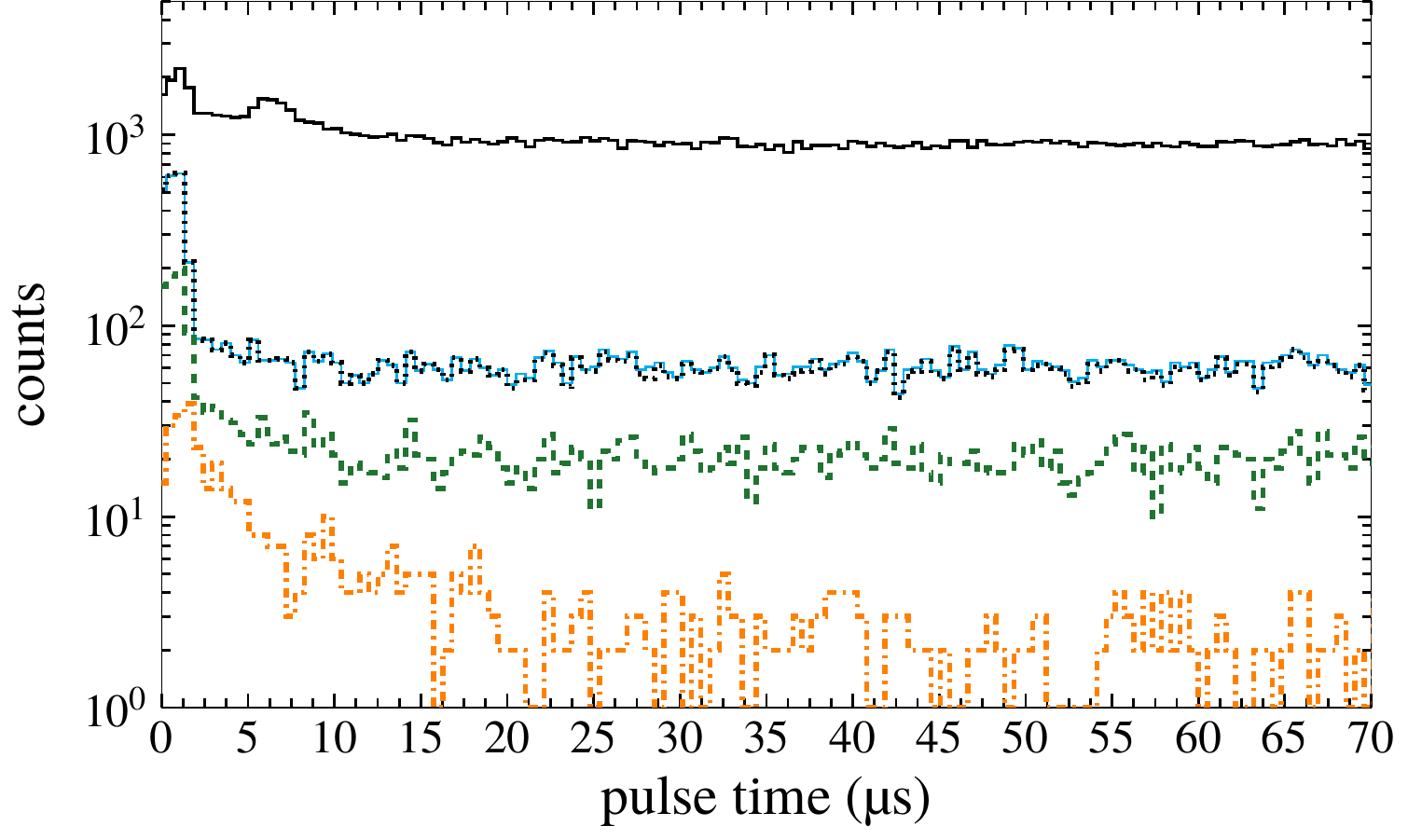}
\caption{Successive applications of each of the neutron candidate event criteria
for beam data taken at position O (center of the tank).
\textbf{SOLID BLACK}: Time distribution of all pulses recorded in DAQ \modeb\
on NCV PMT \#1 at position O. A time of zero corresponds to beam arrival. No analysis cuts have been applied to
these data. 
\textcolor{cyan}{\textbf{SOLID CYAN}}: Time distribution of all NCV coincidences from
the same dataset.
\textcolor{black}{\textbf{DOTTED BLACK}}: Events from the blue histogram that passed the
after-pulsing cut. Note that this cut was not applied to events that occurred
in the first \SI{20}{\micro\second} after the start of the beam spill.
\textcolor{MyGreen}{\textbf{DASHED GREEN}}: Events from the red histogram that passed
the total charge cut.
\textcolor{orange}{\textbf{DASH-DOTTED ORANGE}}: Events from the green histogram that
passed the water PMT veto cut.}
\label{fig:all_pulses_ncv2}
\end{center}
\end{figure}

The solid black curve in \cref{fig:all_pulses_ncv2} shows the time distribution
(relative to the start of the beam spill) for all of the reconstructed pulses
found on one NCV PMT (\#1) during beam data taking in \modeb\ at NCV position O
(the center of the tank). Three features of the distribution are immediately
apparent. The first is a dominant flat component composed primarily of dark
pulses with a contribution from cosmic-ray muons. The second is a peak synchronous with beam arrival. This corresponds primarily to beam neutrino interactions in the
tank, with an admixture of beam-induced muons. A later peak, attributable to a combination of fast neutron scatters and after-pulsing, appears roughly \SI{5}{\micro\second} after the first.

Rather than relying on simulations to estimate
efficiencies for individual analysis cuts, we measure the combined efficiency from all cuts folded with the acceptance of the NCV, as described in \cref{sec:ncv_efficiency}. Thus any cut efficiencies and purities described in this section are provided solely for heuristic purposes and have not been used directly in the final measurement.

\subsection{NCV coincidence requirement}
\label{sec:ncvcoincidence}

To reduce the number of spurious events due to dark noise and after-pulsing, neutron capture candidates were selected by requiring two pulses, one from each NCV PMT, whose start times (calculated as described in
\cref{sec:PMT_pulses}) fell within a
\SI[number-unit-product=\text{-}]{40}{\nano\second} coincidence window. The earlier start time is
designated as the time of the event. The solid cyan curve in
\cref{fig:all_pulses_ncv2} shows the time distribution of neutron candidates remaining after
the NCV coincidence requirement is applied. While the beam-correlated peak at $t
= \SI{0}{\micro\second}$ is still present, the second peak at
\SI{5}{\micro\second} has disappeared, as would be expected if it is largely
composed of after-pulses on NCV PMT \#1.

Given the measured dark rates of the NCV PMTs (\SI{476}{\hertz} for PMT \#1 and \SI{87}{\hertz} for PMT \#2) we
estimate the rate of accidental coincidences for our chosen \SI[number-unit-product=\text{-}]{40}{\nano\second}
coincidence window to be \SI{1.7e-3}{\hertz}. This translates to a negligible
contribution of \num{1.4e-7} counts per spill from accidental NCV
coincidences.

\subsection{After-pulsing requirement}
\label{sec:afterpulse_cut}

NCV coincidences due to after-pulsing can be suppressed by requiring that a coincidence occur at least \SI{10}{\micro\second} after the most recently-accepted neutron candidate. However, it is not possible to distinguish spurious candidates due to after-pulsing and events in which a fast neutron scatters within the NCV shortly before a true neutron capture.
Neutrons produced by trace radioisotopes and cosmic-ray
spallation constitute a CIT background that was considered
in our calibration of the NCV (see \cref{sec:calib}). Apart from these, the only important source
of fast neutrons that may enter the ANNIE detector is the beam itself. We assume
in our definition of the after-pulsing cut that, by \SI{10}{\micro\second} after
the start of the beam spill, any beam-correlated neutrons that are found inside the
detector have dropped below the NCV detection threshold for proton recoils. 
In order to avoid losing signal while still suppressing the majority of after-pulses, the 
the after-pulsing suppression cut is only applied to NCV coincidences recorded between \SIrange[range-phrase=--, range-units=single]{20}{70}{\micro\second} after beam arrival. 

The dotted black histogram in \cref{fig:all_pulses_ncv2} shows the small
effect of applying the after-pulsing cut to the Position O beam data. Given a low overall
probability of producing neutrons and a low probability of two genuine neutron
captures occurring within \SI{10}{\micro\second}, the signal efficiency for this
cut is estimated to approach \SI{100}{\percent} in the time window of interest.

\subsection{Total charge cut}

To suppress NCV coincidences from cosmic- and beam-induced muons, neutron candidates
were eliminated if their energy deposition in the scintillator exceeded the maximum \SI{9}{\MeV} expected from a fully-contained neutron capture $\gamma$-ray cascade (see
\cref{fig:rat_Edep}). Based on the NCV charge-to-energy
calibration described in \cref{sec:cosmic_calib}, a loose cut of $Q_\text{max} = \SI{150}{\pico\coulomb}$ on the maximum total charge on the two NCV PMTs was adopted. 
This conservative choice, which corresponded to a deposited energy of about \SI{34}{\MeV}
(see \cref{eq:NCV_energy_threshold} and \cref{tab:energy_threshold_fit_results}) minimized signal loss. The dashed green histogram in \cref{fig:all_pulses_ncv2} shows the neutron candidates that remain after applying the cut $Q_1 + Q_2 < Q_\text{max}$, where $Q_1$ and $Q_2$ are the charges collected by NCV PMTs \#1 and \#2, respectively, to the candidates in the dotted black histogram. Since this cut is more than three times the energy expected from neutron captures, we expect negligible signal loss.

\subsection{Water PMT veto cut}
\label{sec:waterPMTcut}

\begin{figure}
\begin{center}
\includegraphics{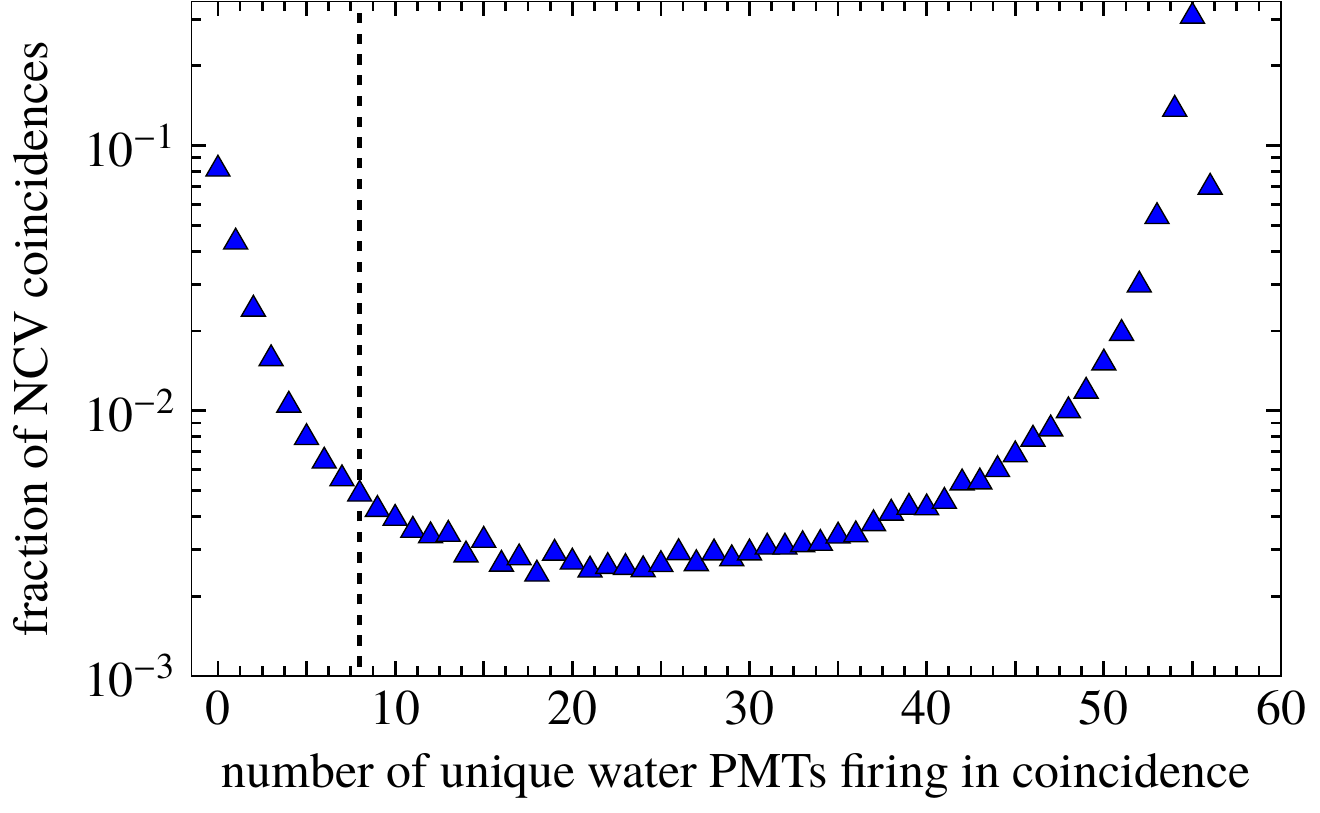}
\caption{Distributions of the number of unique water tank PMTs that recorded at
least one pulse within \SI{40}{\nano\second} of an NCV coincidence. The data
shown here include all runs analyzed for this paper.
Events to the right of the dashed vertical line are removed
by the water PMT veto cut.}
\label{fig:unique_water_pmt_dist}
\end{center}
\end{figure}

Muons that exit the NCV after traveling only a short distance through the scintillator may deposit an energy low enough to pass the NCV total charge cut.
The majority of these muons will produce enough light to activate the PMTs at the bottom of the tank. On the other hand, from simulations we calculate that 98\% of neutron captures occurring within the NCV produce pulses on 8 or fewer water tank PMTs. \Cref{fig:unique_water_pmt_dist} shows the number of water tank PMTs that recorded a pulse within \SI{40}{\nano\second} of an NCV coincidence event (blue
triangles).
The coincident events show a bimodal distribution in the number of tank PMTs that fired. The peak near zero, corresponding to true neutron captures, drops to within a factor of three of the accidental background at 8 PMTs. The peak near 55 PMTs corresponds to muons traversing the NCV. We veto the latter events by
rejecting neutron candidates with more than 8 water tank PMTs firing within
\SI{40}{\nano\second} of the event start time. The resulting pulse time
distribution (after applying this and previous cuts) is shown by the dash-dotted orange
curve in \cref{fig:all_pulses_ncv2}.

This cut removes \SI{99.8}{\percent} of the cosmic-ray calibration sample. Any residual contamination is  addressed via the CIT background subtraction in \cref{sec:results}.

\section{Calibration measurements}
\label{sec:calib}

The calibration measurements required to estimate the neutron detection efficiency of the NCV include a direct measurement of neutrons from a \isotope[252]{Cf} fission source and a measurement of the energy threshold of the NCV.

\subsection{Arrival time distribution of detected neutrons from a
\isotope[\text{252}]{\text{Cf}} source} \label{sec:sourcecal}

Californium-252 is a commonly used radioisotope with a half-life of 2.6
years. In \SI{3.1}{\percent} of its decays, \isotope[252]{Cf} undergoes spontaneous fission to produce an average of \num[group-digits=false]{3.7675(40)} neutrons \cite{TabRad_v4} and \num{7.98(20)} $\gamma$-rays \cite{Valentine2001} per fission. Since the fission $\gamma$-rays and neutrons are emitted nearly simultaneously, a $\gamma$-ray-based trigger provides a clean neutron sample. For our \isotope[252]{Cf} source calibration runs, the NCV was
placed at position V4 at the top center of the tank. A small dark box containing an LYSO \cite{lyso} scintillation crystal coupled to a small photomultiplier tube was placed on the tank hatch above the NCV.
The output of the PMT observing the LYSO crystal was connected to a discriminator, and the \isotope[252]{Cf} source was placed directly above the crystal. Pulses above the discriminator threshold, attributable to fission $\gamma$-rays scattering within the crystal, triggered the acquisition of a \SI[number-unit-product=\text{-}]{80}{\micro\second} DAQ \modea\ readout window with a reduced (\SI[number-unit-product=\text{-}]{2}{\micro\second}) pre-trigger region.

In order to obtain a useful calibration of the neutron detection efficiency, the CIT background and neutron candidates faked by $\gamma$-rays must be separated from the \isotope[252]{Cf} fission neutrons. We do this by measuring the time spectrum of all neutron candidates in the \isotope[252]{Cf} runs relative to the start of their respective data acquisition window.
The structure of the resulting time distribution is shown in
\cref{fig:cf_source_best_fit}. The prompt $\gamma$-rays appear as a sharp
spike in the third bin. The broad bump peaking just before \SI{10}{\micro\second} has an exponential tail whose time constant matches
the expected value for thermal neutron captures in the NCV liquid.
In \cref{sec:ncv_eff1} we fit simulations-derived models for these components to the neutron candidate time distribution to extract the NCV efficiency.

\subsection{NCV energy threshold measurement using cosmic muon data}

The cosmic muon trigger for ANNIE \phasei\ selected a specific
set of downward going muon tracks passing nearly directly through the NCV. This
sample of through-going muons, combined with information from normal beam data,
was used to calibrate the energy threshold of the NCV. This was done in several
steps: \begin{enumerate*}[label=(\arabic*)] \item estimating the charge
threshold for NCV coincidence events from PMT pulse data, \item using through-going
cosmic muon data to estimate the peak charge, and \item combining the peak
charge measurement with information from simulations to obtain a conversion
factor between the summed charge on the two NCV PMTs and the energy deposited in
the NCV liquid scintillator. \end{enumerate*}

\subsubsection{Charge threshold measurement}

Because the algorithms used to reconstruct the PMT pulses rely on a threshold
based on pulse amplitude rather than charge, the total charge collected by the
two NCV PMTs for events at threshold should be distributed about some mean value ($Q_\text{thresh}$). To estimate the mean $Q_\text{thresh}$ of the threshold charge distribution, a Gaussian fit was performed in the vicinity of the peak of the distribution of the total charge ($Q_\text{sum}$) collected on the two NCV PMTs for a large sample
of NCV coincidence events (at all positions) with $Q_\text{sum} <
\SI{100}{\pico\coulomb}$ and with the after-pulsing cut applied. The result of the fit, $Q_\text{thresh} = \SI{20.9}{\pico\coulomb}$, is given in the first row of \cref{tab:energy_threshold_fit_results}.

\subsubsection{Peak charge measurement}
\label{sec:peak_charge_measurement}

To determine the total charge ($Q_{\mu,\text{peak}}$) on the two NCV PMTs associated with the corresponding peak in the cosmic trigger data, a sample of \num{4841} NCV coincidence events recorded at position O (center of the tank) was analyzed.
Each of the selected events occurred within \SI{2}{\micro\second} of a downward
muon candidate being observed by the cosmic ray trigger.
A Gaussian fit to the peak of the cosmic muon charge distribution in the
data yielded the results shown in the second row of
\cref{tab:energy_threshold_fit_results}.

\subsubsection{Calculation of the energy threshold}

To estimate the charge-to-energy conversion factor for the NCV, simulations of
muons passing through the NCV were compared with data taken using the selection from \cref{sec:peak_charge_measurement}. In the simulations, a cosmic muon event generator originally written for the G4beamline code \cite{Roberts2007} was adapted for use with the RAT-PAC detector simulation package \cite{ratpac}.
A Gaussian fit was used to estimate the peak location of the deposited energy
($E_{\mu,\text{peak}}$) at \SI{91.1(2)}{\MeV}.

\begin{table}
\centering
\caption{Results of the fits used to estimate an energy threshold for the
NCV}
\begin{minipage}{\textwidth}
\centering
\renewcommand{\footnoterule}{\vspace{-8pt}}
\begin{tabular}{ccccc}
\toprule
Parameter & Variable &\multicolumn{1}{c}{Best-fit
value\footnote{Parameter errors are statistical only}}
& Fit $\chi^2$ & DOF\footnote{Degrees of freedom} \\
\midrule \addlinespace[6pt]
\multirow{2}{*}{
\makecell[c]{Threshold summed \\ NCV PMT charge peak}}
& \multirow{2}{*}{ $Q_\text{thresh}$ }
& \multirow{2}{*}{ \SI{20.9(3)}{\pico\coulomb} }
& \multirow{2}{*}{ 5.1 } & \multirow{2}{*}{ 14 }
\\
\\
\multirow{2}{*}{
\makecell[c]{Downward muon summed \\ NCV PMT charge peak}} &
\multirow{2}{*}{ $Q_{\mu,\text{peak}}$ }
& \multirow{2}{*}{ \SI{400(8)}{\pico\coulomb} }
& \multirow{2}{*}{ 7.7 } & \multirow{2}{*}{ 7 } \\
\\
\\
\multirow{2}{*}{
\makecell[c]{Simulated downward muon \\ energy deposition peak}}
& \multirow{2}{*}{ $E_{\mu,\text{peak}}$ }
& \multirow{2}{*}{ \SI{91.1(2)}{\MeV} } & \multirow{2}{*}{2.3}
& \multirow{2}{*}{ 6 } \\
\\
\bottomrule
\end{tabular}
\label{tab:energy_threshold_fit_results}
\end{minipage}
\end{table}

Assuming that the total charge measured by the NCV PMTs is approximately a
linear function of the energy deposited in the liquid scintillator, the NCV
energy threshold $E_\text{thresh}$ may be written in the form
\begin{equation}
\label{eq:NCV_energy_threshold}
E_\text{thresh} = Q_\text{thresh} \, \frac{ E_{\mu,\text{peak}} }
{ Q_{\mu,\text{peak}} }
\end{equation}
where $Q_\text{thresh}$ is the summed charge on the NCV PMTs at threshold, and
the ratio $E_{\mu,\text{peak}} / Q_{\mu,\text{peak}}$ is used as a
charge-to-energy conversion factor.

Plugging the best-fit parameter values from
\cref{tab:energy_threshold_fit_results} into \cref{eq:NCV_energy_threshold}
yields the NCV energy threshold $E_\text{thresh} =
\SI[separate-uncertainty=true, parse-numbers=false] {4.76 \pm
0.12_\text{stat}}{\MeV}.$ The statistical error given here was propagated
analytically from the fit results assuming that all three parameters are
independent.

\begin{comment}
\begin{equation}
\label{eq:NCV_energy_threshold_2}
E_\text{thresh} =
\SI[separate-uncertainty=true, parse-numbers=false]
{4.76 \pm 0.12\,(stat)}{\MeV}.
\end{equation}
\end{comment}

\section{Estimation of the NCV efficiency}\label{sec:ncv_efficiency}

The NCV efficiency (i.e., the fraction of true neutron captures within the NCV that are
actually detected) is estimated using two independent techniques. The first technique
relies on a direct measurement of detected neutrons from the \isotope[252]{Cf}
fission source. The second is an indirect estimate based on the energy threshold
of the NCV, as calibrated using through-going cosmic muons. Both
methods rely on simulations to relate the measured quantities to the NCV
efficiency. However, they rely on independent calibration datasets and are primarily sensitive to different aspects of the simulation models. We combine the results from the two approaches to obtain the final measured value of the NCV efficiency and its estimated uncertainty (see \Cref{sec:combinedNCVeffic}).

\begin{figure}
\begin{center}
\includegraphics{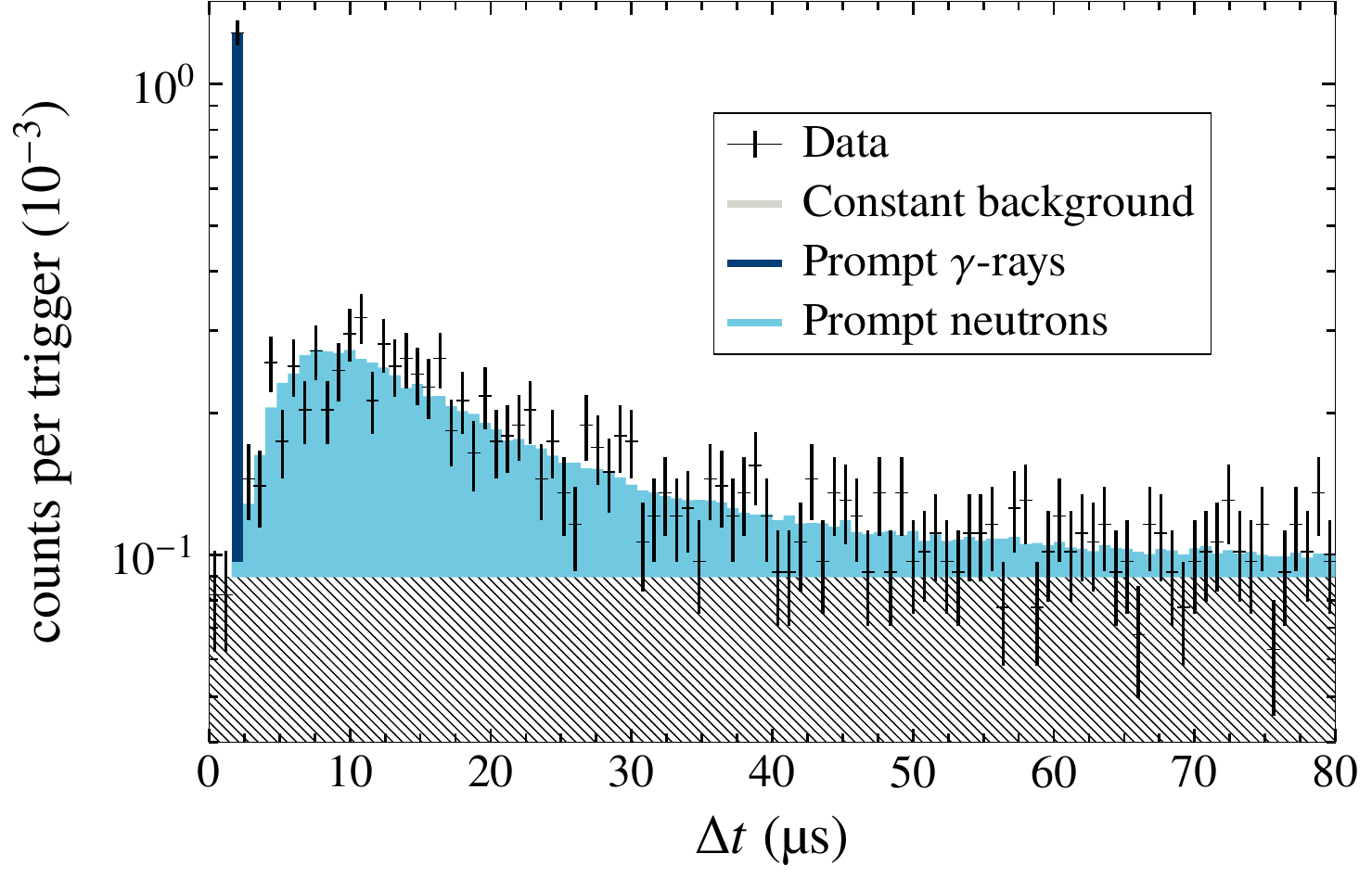}
\caption{Comparison of the \isotope[252]{Cf} calibration source data with the
result of a maximum likelihood fit of the model defined in
\cref{eq:source_model_1}. The fitted contributions of the
constant-in-time background, prompt fission $\gamma$-rays, and prompt neutrons
from the source are shown separately. The horizontal axis gives the time difference between the start of the DAQ \modea\ acquisition window and each neutron candidate event.}
\label{fig:cf_source_best_fit}
\end{center}
\end{figure}

\subsection{Measuring the NCV efficiency with a \isotope[\text{252}]{\text{Cf}}
source}\label{sec:ncv_eff1}

As noted in \cref{sec:data} and \cref{sec:sourcecal}, six \isotope[252]{Cf} source calibration runs were performed with the NCV at position V4, providing a total of \num{206732} \modea\ data acquisition windows taken when the calibration system triggered on a prompt $\gamma$-ray from a \isotope[252]{Cf} fission. \Cref{fig:cf_source_best_fit} shows the arrival time
distribution of neutron candidate events relative to the start of the associated data acquisition window.
This distribution includes three components: a prompt flash from fission $\gamma$-rays interacting in the NCV (shown in dar blue), a flat, CIT component (shown in gray), and an excess following the $\gamma$ flash with a characteristic shape due to neutron captures (shown in light blue).

To extract the NCV efficiency, a maximum likelihood fit to the time distribution in  \cref{fig:cf_source_best_fit} is performed using the ROOT \cite{Brun1997} interface to MINUIT \cite{James:2004xla}.
The fission $\gamma$-rays only contribute to a single time bin and are modeled as a single-bin delta function.  The CIT component is assumed to be flat, which is consistent with the pre-flash region of the time spectrum. The shape of the neutron time distribution is taken from ANNIE RAT-PAC detector simulations together with version 2.0.3 of the FREYA event generator \cite{Verbeke2018,Verbeke2016a}. The neutron cross sections used in these simulations were taken from version 4.5 of the Geant4 Neutron Data Library
\cite{Geant4}.

Formally, the log-likelihood is
\begin{equation}
\ln{\cal{L}} = \sum_j d_j \ln{f_j} - f_j
\end{equation}
with $d_j$ corresponding to the measured number of events in the $j$th time bin and the expected number of counts in the $j$th time bin, $f_j$, given by
\begin{equation}
\label{eq:source_model_1}f_j(\epsilon_{\text{NCV}}, R, P_\gamma) =
N_\text{windows} \left( \epsilon_{\text{NCV}} \, \alpha_{n,j} +
\delta_{j,\gamma\text{flash}} \, P_\gamma
+ \Delta t_j \, R \right)
\end{equation}
The three fit parameters are the NCV efficiency $\epsilon_{\text{NCV}}$, the CIT background rate in Hz ($R$), and the fraction of fissions that result in a $\gamma$-ray detection in the NCV ($P_\gamma$).
The quantity $\alpha_{n,j}$ represents the neutron acceptance of the NCV, i.e., the probability that a \isotope[252]{Cf} fission produces a true neutron capture inside the NCV during the $j$th time bin. This is derived from the
\isotope[252]{Cf} source simulations via the formula
\begin{equation}
\alpha_{n,j} = \frac{N_j} {N_\text{simulated}},
\end{equation}
where $N_j$ is the number of simulated captures that occurred in the $j$th time
bin, and $N_\text{simulated} = \num{e6}$ is the number of simulated fissions.
$N_\text{windows}$ and $\Delta t_j$ are the number of data acquisition windows and the width of a single time bin.

 The values of the best-fit parameters are summarized in \cref{tab:source_fit_results}.
 The comparison of this fit result to the source calibration data is shown in \cref{fig:cf_source_best_fit}.
This method results in a measured NCV efficiency of
\begin{equation} \label{eq:cf_eff} \epsilon_\text{NCV} =
\SI[separate-uncertainty=true, parse-numbers=false] {9.60 \pm
0.57_\text{stat}}{\percent}. \end{equation}

\begin{table}
\centering
\caption{Results of the maximum likelihood fit to the \isotope[252]{Cf} source calibration data. The uncertainties are statistical.}
\medskip
\begin{tabular}{ccS[table-format=1.44(22)e2,table-alignment=center]}
\toprule
Parameter & Variable &\multicolumn{1}{c}{Best-fit value} \\
\midrule
NCV efficiency & $\epsilon_\text{NCV}$ & \num{9.60(57)e-2} \\
Background event rate (Hz) & $R$ & \num{1.12(4)e2} \\
$\gamma$-ray event probability & $P_\gamma$ & \num{1.19(8)e-3} \\
Reduced chi-squared statistic & $\chi^2 / \nu$ &  \num{0.753} \\
\bottomrule
\end{tabular}
\label{tab:source_fit_results}
\end{table}

\subsection{Estimating the NCV efficiency using the energy threshold}
\label{sec:cosmic_calib}

The second method for estimating the NCV efficiency uses simulations to predict the fraction of true neutron captures that deposit energy in the NCV liquid scintillator above the measured detection threshold of 4.76 MeV. The black curve in \cref{fig:rat_Edep} shows the distribution of the total energy deposition within the scintillator for the \num{70470} simulated neutron captures that occured in the NCV liquid volume. A negligible number of external neutron captures produced energy deposits in the NCV.
The dashed blue line in \cref{fig:rat_Edep} shows the measured energy threshold of the
NCV.

The NCV efficiency, $\epsilon_\text{NCV}$  is the ratio of the simulated NCV capture events with energy depositions above $E_\text{thresh}$, divided by the total number of simulated NCV captures:

\begin{figure}
\begin{center}
\includegraphics{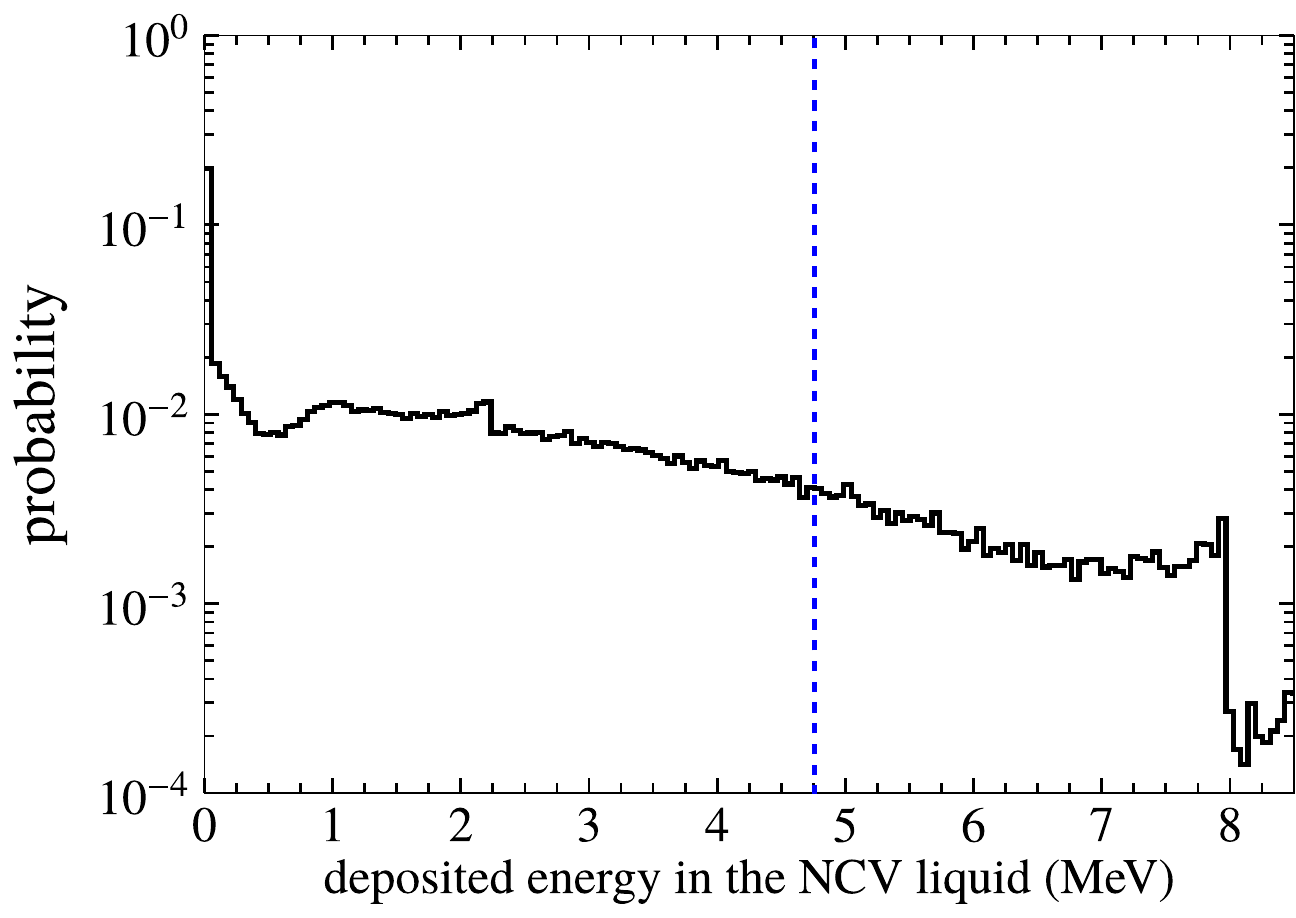}
\caption{Total energy deposited in the scintillator for
simulated neutron captures that occurred within the NCV liquid volume.
The estimated NCV energy threshold of
\SI{4.76}{\MeV} is indicated by the dashed blue line.}
\label{fig:rat_Edep}
\end{center}
\end{figure}

\begin{equation}
\label{eq:rat_ncv_eff_value}
\epsilon_\text{NCV} =
\SI[separate-uncertainty=true, parse-numbers=false]
{12.8 \pm 0.9\,(stat)}{\percent}.
\end{equation}
The statistical error shown in \cref{eq:rat_ncv_eff_value} was found
by computing $\epsilon_\text{NCV}$ with the value of $E_\text{thresh}$ adjusted
by a plus or minus one-sigma error.

\section{Computing the beam-induced neutron event count}
\label{sec:results}

\subsection{Subtraction of the constant-in-time (CIT) background}
\label{sec:CIT_subtraction}

As discussed in \cref{sec:intro}, neutron backgrounds for ANNIE's \phaseii\
physics measurements will consist of a CIT component arising
from natural radioactivity and a component correlated in time with spills from
the BNB. Because the first of these components can be characterized {\it in situ} using an off-beam or zero-bias trigger, the goal of ANNIE \phasei\ is to isolate and estimate the rate of beam-induced neutron backgrounds in the detector. 

We obtain an estimate of the number of beam-induced neutron events at each NCV position by subtracting an estimate of the CIT component from the total count of neutron candidates observed within \SIrange[range-phrase=--, range-units=single]{10}
{70}{\micro\second} window after the start of each beam spill.

\subsubsection{CIT background estimation using pre-beam data}
\label{sec:cit_pre_beam}

All of the data collected in position V4 and a portion of the position O data were recorded in DAQ \modea, where the acquisition time window included \SI{10}{\micro\second} of pre-beam data. In order to ensure that the sample had no contamination from the beam,  the \SI{1}{\micro\second} prior to beam start is excluded, leaving a total of \SI{9}{\micro\second} per trigger for estimating the CIT background. We therefore designate the number of events from the first \SI{9}{\micro\second} of the DAQ \modea\ readout window that pass all selection cuts as $N_\text{pass}^\text{pre}$. \Cref{fig:NCVtimes} shows that the pre-beam event rate is
substantially higher at position V4 than with the considerable shielding at position O.
For those positions where \modea\ data are not available, we used the
pre-beam data from the most shielded position (O) to estimate the CIT
background.

The general equation for $N_n^\text{pre}$, the pre-beam estimate for the
number of neutron candidates attributable to CIT background, is
\begin{equation}
\label{eq:CIT1}
N_n^\text{pre} = \frac{ \Delta t^\text{ROI} }{ \Delta t^\text{pre}} N_{pass}^\text{pre}\frac{ \mathcal{T} }{
\mathcal{T}_\text{O} }.
\end{equation}
Here the scaling factor $\frac{ \Delta t^\text{ROI} }{ \Delta t^\text{pre}}$
accounts for the difference in duration of the post-beam region of interest ($\Delta
t^\text{ROI} = \SI{60}{\micro\second}$) and the pre-beam region used to estimate the CIT background ($\Delta
t^\text{pre} = \SI{9}{\micro\second} $). The measured systematic uncertainty in the
time intervals $\Delta t^\text{ROI}$ and $\Delta t^\text{pre}$ is less than one part in $10^5$ and is therefore neglected in this analysis.
The factor $\mathcal{T}/\mathcal{T}_\text{O}$ is the ratio of recorded
beam triggers for the position in question and position O, and is only applicable for the positions where \modea\ data are not available.

\subsubsection{CIT background estimation using late-time data}
\label{sec:cit_late_time}

At some point after a beam trigger (but before the arrival of a new beam trigger) the event rate should return to baseline. For positions with \modeb\/ data available we had access to neutron candidate events recorded after the signal region of interest and used a period \SIrange[range-phrase=--, range-units=single]{70}{80}{\micro\second} after beam arrival to obtain a second, independent estimate of the CIT component. We denote the number
of events in this time period that pass all selection cuts by $N_\text{pass}^\text{post}$. The late-time estimate of the number of counts
attributable to the CIT background can then be written as
\begin{equation}
\label{eq:CIT2}
N_n^\text{post} = \frac{ \Delta
t^\text{ROI} }{ \Delta t^\text{post}} N_{pass}^\text{post}
\end{equation}
where $\Delta t^\text{post} = \SI{10}{\micro\second}$.

\subsection{After-pulsing correction}
\label{sec:apcorr}

As previously noted, the after-pulsing cut is applied uniformly to the pre-beam
data but is not applied to neutron candidate pairs in beam data when the first
neutron arrives during the initial \SI{10}{\micro\second} after the beam. This
prevents the suppression of neutron captures that follow shortly after proton
recoils induced by fast neutrons. However, this also means that true
after-pulses are not suppressed during this same time period.

\begin{table}
\centering
\caption{Measurements used to obtain an estimate of the ratio of
after-pulses to neutron candidate events $R_\text{after-pulse}$.}
\medskip
\begin{tabular}{cS[table-alignment=center]S[table-alignment=center]S[table-format=1.44(22)e2,table-alignment=center]}
\toprule
Signal criteria & {Signal events} & {After-pulses}
& {$R_\text{after-pulse}$ (\si{\percent})} \\
\midrule
Beam data, mode A, position V4 & 2464 & 10 & \num{0.41(12)} \\
Beam data, mode B, all positions & 1567 & 13 & \num{0.83(21)} \\
Cf source data, mode A, position V4 & 1162 & 8 & \num{0.69(22)} \\
\midrule
Weighted mean & & & \num{0.54(09)} \\
\bottomrule
\end{tabular}
\label{tab:afterpulses}
\end{table}
In order to correct for this effect, we calculate the ratio of after-pulses to
neutron candidate events by comparing the neutron candidate yield before and
after applying the after-pulsing suppression cut. This is done for
\SI[number-unit-product=\text{-}]{40}{\micro\second} time periods starting \SI{20}{\micro\second} post-trigger,
using both \isotope[\text{252}]{\text{Cf}} calibration data and beam data taken
at several positions. The resulting after-pulse rates, with statistical errors, are summarized in  \cref{tab:afterpulses}.
After-pulse-per-neutron rates obtained in the first
\SIrange{10}{20}{\micro\second} after beam crossing are roughly a factor of two
higher, confirming the presence of additional fast neutrons during this time
period.
We take a weighted average of all results in \cref{tab:afterpulses} to obtain
an after-pulse-per-neutron rate of $R_{\text{after-pulse}}= 0.54 \pm 0.09_\text{stat} \si{\percent} $.

\subsection{Beam-correlated neutron event counts}
\label{sec:nEventCounts}

The final number of beam-correlated neutron candidate events, corrected for both the CIT background and after-pulsing, is given at any position
by
\begin{equation}
\label{eq:total_neutron_counts_background_subtracted}
N_n = (1 - R_{\text{after-pulse}})N_n^\text{\SI{10}{\micro\second}}
+ N_n^\text{later} - N_n^\text{CIT}
\end{equation}
where $N_n^\text{\SI{10}{\micro\second}}$ ($N_n^\text{later}$) is
the raw neutron count in the first \SI{10}{\micro\second} (remaining \SI{50}{\micro\second}) of our signal region of interest. The quantity
\begin{equation}
\label{eq:weighted_mean_CIT}
N_n^\text{CIT} = \frac{ w_\text{pre} \, N_n^\text{pre}
+ w_\text{post} \, N_n^\text{post} }
{ w_\text{pre} + w_\text{post}}
\end{equation}
is a weighted average of the two different CIT background estimates where the weights
\begin{align}
w_\text{pre} &\equiv
\Big( N_{pass}^\text{pre} \Big)^{-1}
\Big( \frac{ \Delta t^\text{ROI} }{ \Delta t^\text{pre}}
\frac{ \mathcal{T} }{ \mathcal{T}_\text{O} } \Big)^{-2}
\\[0.5\baselineskip]
w_\text{post} &\equiv
\Big( N_{pass}^\text{post} \Big)^{-1}
\Big( \frac{ \Delta
t^\text{ROI} }{ \Delta t^\text{post}} \Big)^{-2}
\end{align}
are the reciprocals of the statisical variances from each measurement. The neutron event count results are summarized in \cref{tab:neutron_count_results}. In the third column, the statistical uncertainty on $N_n^\text{CIT}$ is given by the standard error
$(w_\text{pre} + w_\text{post})^{-1/2}$.
At position V4, for which no \modeb\ data were taken, we use $N_n^\text{CIT} = N_n^\text{pre}$ with a statistical uncertainty of $w_\text{pre}^{-1/2}$.

\section{Computing the beam-induced neutron event rate}
\label{sec:rate}

\begin{figure}
\begin{center}
\includegraphics{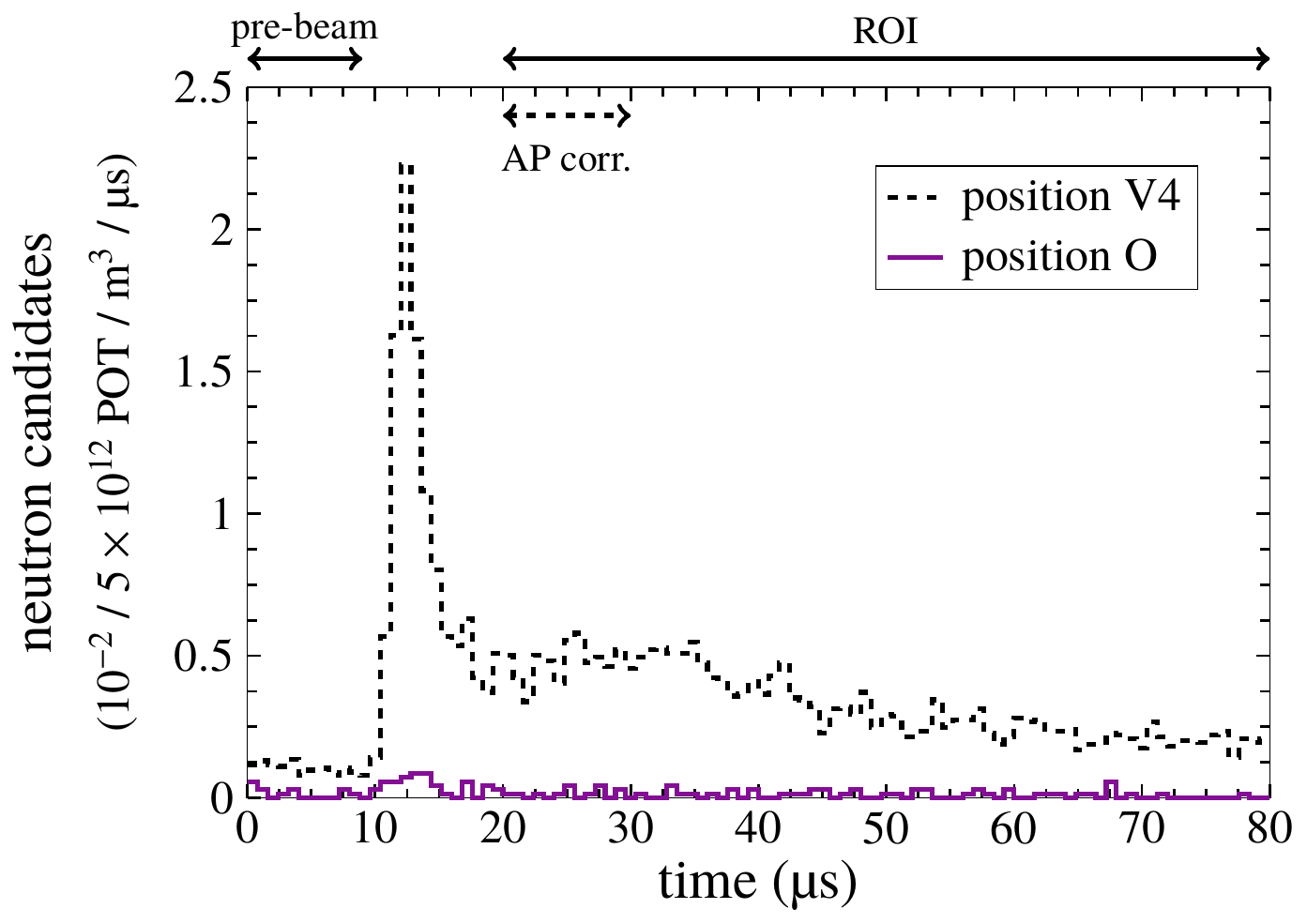}
\caption{Time distributions of the neutron event rates measured at NCV position
V4 (top of the tank, dashed black) and at position O (center of the tank, solid purple). In
the DAQ \modea\ data shown here, the neutrino beam arrives at
\SI{10}{\micro\second}. At the top of the plot, the time intervals used to measure beam-correlated neutron captures (``ROI'') and the constant-in-time background (``pre-beam'') are indicated with solid lines. A dashed interval (``AP corr.'') is also used to mark the early portion of the ROI in which a correction for after-pulsing is applied in the analysis (see \cref{sec:apcorr}). }
\label{fig:NCVtimes}
\end{center}
\end{figure}

\Cref{fig:NCVtimes} shows the exposure-normalized neutron candidate event rates
as a function of time for NCV positions V4 and O, recorded using DAQ
\modea. Both distributions show a peak in
coincidence with the neutrino beam due to prompt activity, distinct from
neutron captures. In the center of the tank, a large fraction of the prompt
activity likely corresponds to neutrino interactions within the NCV. At the
surface, this activity extends a few microseconds after the beam and is likely
dominated by fast neutrons scattering off of nuclei in the scintillator.

\begin{table}
\centering
\caption{Beam-correlated neutron event rate measurements performed at each NCV position. Columns from left to right: the number of neutron candidate events observed in the time region of interest ($N_n^\text{ROI} \equiv N_n^{\SI{10}{\micro\second}} + N_n^\text{later}$), the estimated number of events ($N_n^\text{CIT}$) attributable to constant-in-time (CIT) background, the number of events after correcting for the CIT background and after-pulsing ($N_n$), and the beam-correlated neutron event rate $\mathcal{R}_n^\text{NCV}$ per unit volume per beam spill.}
\begin{adjustbox}{center=\textwidth}
\centering
\small
\def\arraystretch{1.4}
\begin{tabular}{cc
  S[table-format=4.0] @{$\,\pm\,$} S[table-format=2.0] @{$\,_\text{stat}\,\pm$} l
  S[table-format=4.0] @{$\,\pm$} l
  r @{$\,\pm\,$} r @{$\,\pm\,$} l
  }
\toprule
{\multirow{2}{*}{
\makecell[b]{NCV \\ position}}} &
& \multicolumn{5}{c}{} & \multicolumn{3}{c}{{$\mathcal{R}_n^\text{NCV}$}} \\
& {$N_n^\text{ROI}$} & \multicolumn{3}{c}{{$N_n^\text{CIT}$}}
& \multicolumn{2}{c}{{$N_n$}} &
\multicolumn{3}{c}{ (\si{\percent\per\meter\cubed\per\spill}) } \\
\midrule
O & 339
& 333 & 45 & { 69$\,_\text{syst}$ }
& 5 & { 48$\,_\text{stat}$ }
& 0.013 & { 0.11$\,_\text{stat}$ }
& { 0.16$\,_\text{syst}$ } \\ \hline
H1 & 60
& 41 & 11 & { 21$\,_\text{syst}$ }
& 19 & { 13$\,_\text{stat}$ }
& 0.35 & { 0.24$\,_\text{stat}$ }
& { 0.40$\,_\text{syst}$ } \\
H2 & 743
& 609 & 56 & { 192$\,_\text{syst}$ }
& 133 & { 62$\,_\text{stat}$ }
& 0.41 & { 0.19$\,_\text{stat}$ }
& { 0.60$\,_\text{syst}$ } \\ \hline
V1 & 254
& 206 & 30 & { 22$\,_\text{syst}$ }
& 47 & { 34$\,_\text{stat}$ }
& 0.29 & { 0.20$\,_\text{stat}$ }
& { 0.15$\,_\text{syst}$ } \\
V2 & 866
& 540 & 51 & { 229$\,_\text{syst}$ }
& 325 & { 59$\,_\text{stat}$ }
& 1.2 & { 0.23$\,_\text{stat}$ }
& { 0.9$\,_\text{syst}$ } \\
V3 & 368
& 140 & 22 & { 124$\,_\text{syst}$ }
& 227 & { 29$\,_\text{stat}$ }
& 2.6 & { 0.35$\,_\text{stat}$ }
& { 1.5$\,_\text{syst}$ } \\
V4 & 3825
& 1207 & 90 & { 0$\,_\text{syst}$ }
& 2613 & { 109$\,_\text{stat}$ }
& 13.6 & { 0.9$\,_\text{stat}$ }
& { 3.1$\,_\text{syst}$ } \\
\bottomrule
\end{tabular}
\end{adjustbox}
\label{tab:neutron_count_results}
\end{table}

After removing the CIT background contribution, we can calculate
the number of beam-induced neutrons that captured in the NCV by
\begin{equation}
\label{eq:N_NCV_2}
\mathcal{N_{{\mathrm n}}^\text{NCV} }= \frac{N_n}
{ \epsilon_\text{NCV} },
\end{equation}
where $N_n$ is the background-subtracted number of observed neutron
candidate events integrated over the time window of interest and
$\epsilon_\text{NCV}$ is the NCV efficiency.

The beam-induced neutron event rate (i.e., neutron captures per unit volume per spill) is calculated at each position from
\begin{equation}
\label{eq:R_NCV_2}
\mathcal{R}^\text{NCV}_n  = \frac{\mathcal{N_{\mathrm n}^\text{NCV} }}
{ \mathcal{P}\, V_\text{NCV} },
\end{equation}
where the exposure $\mathcal{P}$ is the
total number of \si{\pot} normalized to nominal spills of \SI{5e12}{\pot}, and $V_\text{NCV}$ is the volume of the NCV
liquid. The results of this calculation, with full statistical and systematic errors, are summarized in \cref{tab:neutron_count_results} and graphically illustrated in \cref{fig:rates1}. Using these rates and correcting for differences in the time constant and capture rates between 0.25\% Gd-loaded scintillator and 0.1\% Gd-loaded water, it is possible to estimate the expected beam-induced neutron background rates for ANNIE \phaseii\/.

\section{Systematic uncertainties in the beam-induced neutron event rate}

\subsection{Systematic uncertainties on $N_n$}
\label{sec:syst_Nn}

\begin{table}
\centering
\caption{Constant-in-time (CIT) background event rate estimates at each NCV position.
In \modea, the CIT event rate is estimated using the first \SI{9}{\micro\second} of each
readout window (which precedes the arrival of the beam). In \modeb,
the \SI{10}{\micro\second} following our time region of interest
(\SIrange[range-phrase=--, range-units=single]{70}{80}{\micro\second}
after beam arrival) is used.}
\label{tab:CITrates}
\begin{minipage}{\textwidth}
\renewcommand{\footnoterule}{\vspace{-5pt}}
\centering
\def\arraystretch{1.4}
\begin{tabular}{ccr}
\toprule
& {\multirow{2}{*}{\makecell[b]{NCV \\ position}}}
& {\multirow{2}{*}{\makecell[b]{CIT event \\ rate (\si{\hertz})}}} \\
& & \\
\midrule
{\multirow{2}{*}{\makecell[b]{pre-beam \\ (\modea)}}} &
 V4 & \num{11.4(8)} \\
& O & \num{1.5(4)}  \\
\midrule
{\multirow{6}{*}{\makecell[b]{late-time \\ (\modeb)}}} &
   O & \num{1.2(2)}  \\
& H1 & \num{0.8(3)}  \\
& H2 & \num{2.6(2)}  \\
& V1 & \num{1.8(3)}  \\
& V2 & \num{3.1(3)}  \\
& V3 & \num{4.5(7)}  \\
\bottomrule
\end{tabular}
\end{minipage}
\end{table}

The largest systematic uncertainty on the raw neutron count $N_n$ arises from
the CIT background subtraction. We have two independent estimates of the CIT
background at every position except V4 (where only \modea\ data are
available). For positions other than V4 and O, the pre-beam estimate
of the CIT background rate (see \cref{sec:cit_pre_beam}) is taken to be identical to the most shielded
position (O) and is thus likely to be an underestimate. The post-beam method, on the other hand, has the potential to overestimate
the CIT background due to beam contamination. \Cref{tab:CITrates}
summarizes the CIT event rates obtained with both methods. 
We see that for the inner positions (O, V1, and H1) the post-beam CIT background estimate is consistent with the position O pre-beam estimate. For the positions closer to the edge and top of the tank, the differences between the post-beam and position O pre-beam estimate noticeably increase, leading them to dominate the systematic uncertainty.
As described in \cref{sec:nEventCounts}, we estimate the final CIT background count at each position ($N_n^\text{CIT}$) and its statistical
uncertainty using a statistically weighted mean of the two measurements and its standard
error. Treating the two measurements as belonging to a simple random sample (of size two) allows one to compute a sample standard deviation
\begin{equation}
\mathrm{SD}(N_n^\text{CIT}) = \Bigg[
\frac{ w_\text{pre} \, (N_n^\text{pre} - N_n^\text{CIT})^2
+ w_\text{post} \, (N_n^\text{post} - N_n^\text{CIT})^2
}
{w_\text{pre} + w_\text{post}
- (w_\text{pre}^2
+ w_\text{post}^2)
/ (w_\text{pre} + w_\text{post})
}
\Bigg]^{1/2}
\end{equation}
which we take as the systematic error.

The systematic uncertainty on the after-pulsing subtraction is small and has been neglected. Because of this the systematic errors on $N_n$ are identical to those on $N_n^\text{CIT}$ and have been omitted in \cref{tab:neutron_count_results}.

\subsection{Measurement of $\epsilon_\text{NCV}$ and associated systematic error}
\label{sec:combinedNCVeffic}

To determine the final measured value of the NCV efficiency, we adopt the same general approach that was used to combine two independent measurements of the CIT neutron candidate event rate (see \cref{sec:nEventCounts,sec:syst_Nn}). In this case, the two measurements of interest are the NCV efficiency calibrations performed with a \isotope[252]{Cf} source (\cref{sec:ncv_eff1}) and with cosmic-ray muons (\cref{sec:cosmic_calib}). The NCV efficiency and its statistical error are calculated using a statistically weighted mean of the two measurements and its standard error. The reciprocal of each measurement's statistical variance is used as a weighting factor. As in \cref{sec:syst_Nn}, we treat the two efficiency measurements as forming a simple random sample, and we take the sample standard deviation as an estimate of the systematic uncertainty. The combined measurement of the NCV efficiency is thus given by
\begin{equation}
\label{eq:ncv_eff_final}
\epsilon_\text{NCV} =
\SI[separate-uncertainty=true, parse-numbers=false]
{10.5 \pm 0.5_\text{stat} \pm 2.3_\text{syst}}{\percent}.
\end{equation}

\subsection{Systematic uncertainties in exposure and volume scaling}

For the uncertainty on beam exposure $\mathcal{P}$, we adopt the
\SI{2}{\percent} systematic uncertainty found during routine calibrations of
the beam current toroids \cite{MinibooneFlux}. We also adopt uncorrelated
uncertainties of \SI{1.27}{\centi\meter} for the NCV vessel's outer dimensions and \SI{0.16}{\centi\meter} for its wall thickness, leading to a relative uncertainty on the NCV liquid volume $V_\text{NCV}$ of \SI{5.7}{\percent}.

\subsection{Combined uncertainty estimate}

The combined statistical and systematic uncertainty on the beam-induced neutron event rate $\mathcal{R}^\text{NCV}$ at each position is derived by analytically propagating both the statistical and the previously described systematic uncertainties. The uncertainties on the factors $\frac{\Delta t^\text{ROI} }{ \Delta t^\text{pre}}$, $\frac{\Delta t^\text{ROI} }{ \Delta t^\text{post}}$ and $ \frac{ \mathcal{T} }{\mathcal{T}_\text{O} }$ (used to estimate the CIT background) are negligible and therefore omitted. The resulting systematic uncertainties appear in the last column of \cref{tab:neutron_count_results}.

\begin{figure}
\begin{center}
\includegraphics{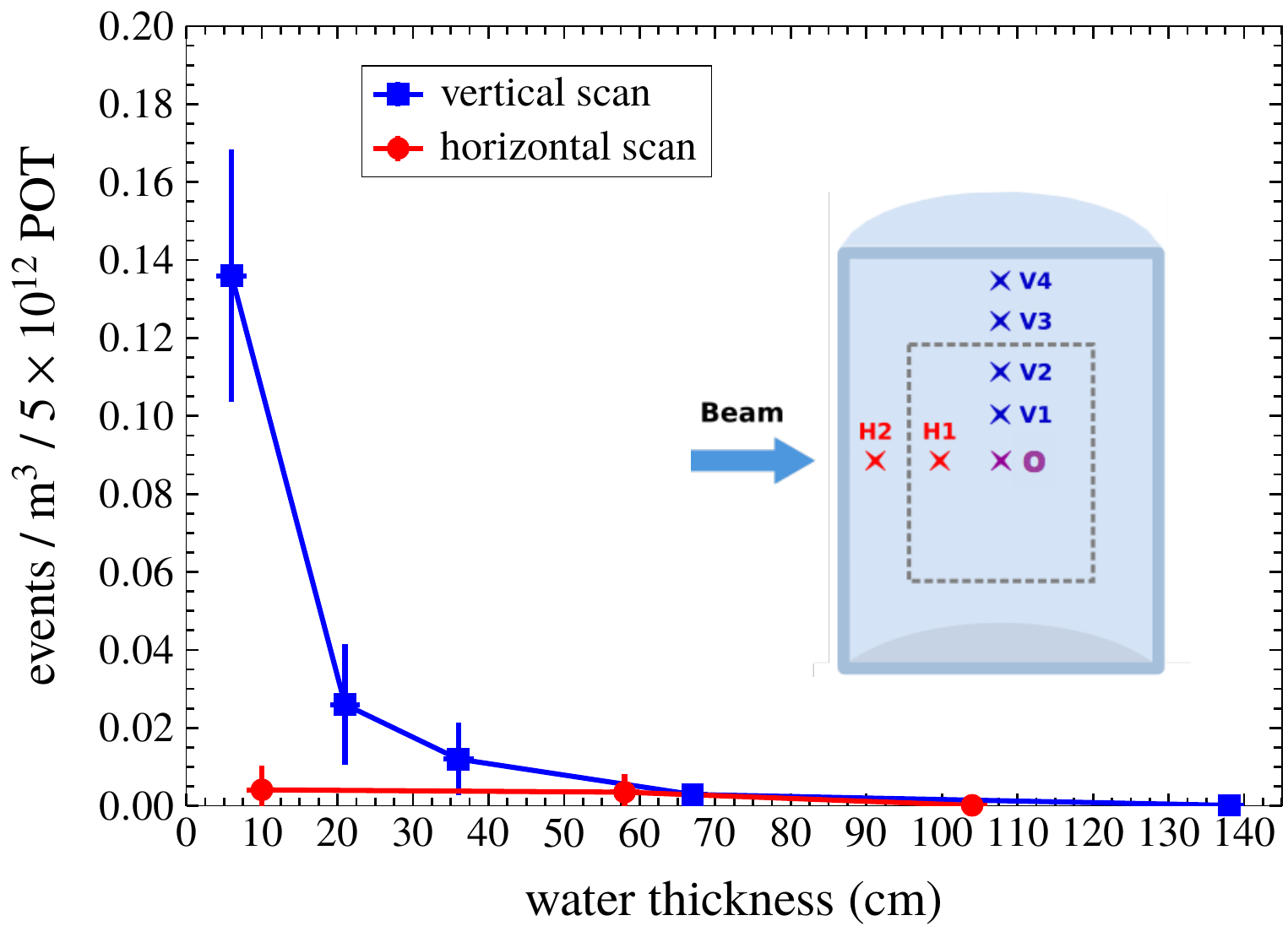}
\caption{Beam-correlated neutron candidate event rates measured during ANNIE \phasei. For the blue square data points, the ``water thickness'' reported on the horizontal axis is the depth of the water above
the top of the NCV. For the red circle data points, it is the smallest distance between the side of the tube forming the NCV vessel and the beam side of the tank. The inset diagram shows the NCV positions included in the red and blue datasets. Position O (the center of the tank) is
shown in purple to indicate that it is included in both the red and blue data. The dashed line indicates which NCV positions are contained within
the active region of ANNIE \phaseii. Error bars shown in the plot include both statistical and systematic contributions.}
\label{fig:rates1}
\end{center}
\end{figure}

\section{Implications for the ANNIE neutron multiplicity measurement}

We can predict both the dominant source and expected rate of beam-correlated background neutrons for the ANNIE~\phaseii\/ physics program. \Cref{tab:neutron_count_results} shows that the background event rate observed at the top center of the tank (i.e., position V4) is significantly larger than at all other positions, including the most upstream location, H2. This result is consistent with the beam-induced neutron
background being dominated by sky-shine rather than dirt neutrons, and it qualitatively agrees with previous SciBooNE results that showed a large excess
of events near the top of
the detector briefly after the beam crossing~\cite{Takeiskyshine}.
The rapid fall-off of the background event rate over just \SI{15}{\centi\meter} of depth is consistent with a soft neutron energy spectrum.

In order to ensure efficient containment of final-state neutrons, the \phaseii\ measurement will select only events with neutrino interaction vertices in a small
($\sim$\SI{2.5}{\meter\cubed}) fiducial region vertically centered in the tank
and slightly upstream of the tank center in the beam direction. The full water
volume will be doped with a \SI{0.1}{\percent} concentration by mass of dissolved Gd. Neutron capture candidates will be accepted anywhere in the
$\sim$\SI{14}{\meter\cubed} active volume bounded by the tank PMTs.

The dashed line on the inset of \cref{fig:rates1} shows which NCV positions are
located within the active detection volume of the ANNIE \phaseii\ detector. The
active region will be located from \SI{36}{\centi\meter} below the water line
(at the top) to \SI{353.5}{\centi\meter} below the water line (at the bottom).
Similarly, the octagonal footprint of the inner region of the detector will be
\SI{20.3}{\centi\meter} away from the wall of the tank at the octagon corners and
as far as \SI{27.2}{\centi\meter} at the midpoint of each side. The highest
beam-induced background neutron rate within this active volume was measured at
position V2 to be 0.012 neutrons per \si{\meter\cubed} per spill. This rate continued to
drop with depth until position O where it is consistent with zero within
errors.

The ANNIE \phaseii\ detector is expected to see an average of no more than one 
neutrino interaction per $\sim$150 BNB spills. Since neutrino interactions and background neutrons are statistically independent, the per-spill neutron rate can be thought of as the probability of detecting a background neutron following a signal neutrino
interaction in \phaseii.

Using the estimated background rates within the expected \SI[number-unit-product=\text{-}]{14}{\cubic\metre} active volume
of ANNIE \phaseii, it is possible to place an upper bound on the contribution of the
beam-induced neutron background to ANNIE signal events. The projections given here are highly conservative and rely on two
assumptions. First, the rates along the horizontal scan are assumed to be
radially symmetric. This assumption is likely accurate for any side-penetrating
neutrons that originate from sky-shine. The dirt-neutron rates would, if
anything, be lower on the downstream side of the tank. Second, we take the
background rates below \SI{138}{\centi\meter} of water to be constant and
consistent with those measured along the horizontal scan at
\SI{138}{\centi\meter} depth (positions O, H1 and H2).

With these assumptions, we integrate the interpolated rates over the \SI[number-unit-product=\text{-}]{14}{\cubic\metre} active volume to obtain an estimated average rate of \tankRateEquation\ beam-correlated background neutrons per ANNIE \phaseii\ signal event. This is nearly an order of magnitude below the expected 0.42 primary neutrons per charged-current neutrino interaction derived from GENIE simulations. 

To account for correlations between NCV positions when assessing the uncertainty on $\mathcal{R}_\text{n}^\text{tank}$, we relied on Monte Carlo simulations. In a set of five hundred thousand trials, $N_n$ for each NCV 
position was varied about its measured value based on the statistical and
systematic uncertainties for each term in
\cref{eq:total_neutron_counts_background_subtracted}. 
With the exception of $N_n^{\SI{10}{\micro\second}}$ and $N_n^\text{later}$, which were treated as Poisson random variables, all other quantities were varied by sampling corrections
from a normal distribution with mean zero and standard deviation equal to the quoted uncertainty of interest. 
The NCV detection
efficiency $\epsilon_\text{NCV}$, total exposure $\mathcal{P}$, and NCV volume
$V_\text{NCV}$ were likewise varied about their central values for each trial,
but the same factors were used at all NCV positions. Unphysical negative rates
were set to zero during each trial in agreement with the ``method of
sensitivity limit'' proposed by Lokhov and Tkachov \cite{Lokhov2015}. One-sigma
errors on $\mathcal{R}_\text{n}^\text{tank}$ were obtained by computing empirical
\SI{68.27}{\percent} confidence intervals using the results from the Monte
Carlo trials.

There are two considerations that will bias our estimate of $\mathcal{R}_\text{n}^\text{tank}$ slightly high relative to the true neutron background in \phaseii\/. The first is that the shielding effect from Gd-loaded water (where thermal neutrons have a shortened diffusion length) is likely to be slightly higher than the measured shielding effect from the pure
water volume in \phasei\/. The second is that the neutron capture time in the Gd-loaded scintillator of the NCV is shorter than that in Gd-loaded water. This means that the \phasei\/ signal window will capture slightly more background neutrons (which have a higher probability of coming in at late times relative
to the beam) relative to the same window in \phaseii\/. We can therefore confidently say that the beam-induced neutron background in ANNIE \phaseii\/ will be acceptably low.

\section{Conclusions}

In this paper we present an estimate of neutron backgrounds derived from
measurements in the Neutron Capture Volume of the ANNIE \phasei\ detector. Quantifying the size of these backgrounds is important in establishing the feasibility of the ANNIE \phaseii\ physics program.

Neutron backgrounds are highest at the
top of the tank at a rate of $0.136 \pm 0.009_\text{stat} \pm 0.031_\text{syst}$ per cubic meter per spill. These backgrounds drop off rapidly with
depth to be consistent with zero for most of the inner volume. With all of our
assumptions erring on the side of overestimating these backgrounds, we still
obtain an event rate of
\tankRateEquation\ 
beam-induced background neutrons per neutrino interaction in ANNIE \phaseii.
Comparing this result with a GENIE prediction of 0.42 neutrons per charged-current neutrino interaction in ANNIE allows us to
conclude that the beam-correlated background neutron rate is acceptably low for the \phaseii\ physics measurements.

The position dependence of these backgrounds is consistent with a flux of low-energy sky-shine neutrons, mostly at the top of the tank, that
drops off rapidly with depth. Optically
isolating the active volume of the ANNIE \phaseii\ detector
\SI{36}{\centi\meter} below the top of the water line and \SI{20}{\centi\meter}
from the side will suffice to reduce these backgrounds to an acceptable rate.

The results presented in this paper are relevant to other BNB experiments such as
SBND, located adjacent to ANNIE Hall,
where dirt neutrons and sky-shine could present similar backgrounds. The
techniques described in this paper will also be applicable to any future
water-based near detectors, especially those with Gd-loading or water-based
liquid scintillators. The operational experience gained during
\phasei\ has informed the design of ANNIE \phaseii.

\section*{Acknowledgements}

This work could not have been accomplished without the support, personnel,
facilities, and resources of Fermi National Accelerator Laboratory. The collaboration owes a special thanks to James Kilmer and the rest of the engineering staff for their guidance in the design and mechanical aspects of the detector. John Voirin and his technical staff were essential in the
deployment and staging of this experiment. Geoffrey Savage and Michael Cherry provided critical guidance and assistance in the process of designing, certifying, and deploying our readout electronics. A special thanks to Yau Wah at the University of Chicago for his generous loan of the K0TO readout boards used in the data acquisition system. The ANNIE collaboration would like to dedicate the success of this effort to the memory of William ``Bill" Lee, our experimental liaison who served as a guide, mentor, and friend to so many of our collaborators. The activities of the ANNIE experiment are supported by the U.S. Department of Energy, Office of Science, Office of High Energy Physics under contracts DE-SC0016326 and DE-SC0019214, together with the Science and Technology Facilities Council and the Scottish Universities Physics Alliance (United Kingdom). JF~Beacom is supported by NSF Grant PHY-1714479. M~Wetstein is supported by DE-SC0017946 and his startup fund at Iowa State. M.~Sanchez, F.~Krenrich, and A.~Weinstein are supported by DE-SC0015684. M. Sanchez would like to acknowledge her support under NSF-1056262 and M. Wetstein acknowledges support through the McCormick Fellowship at U. Chicago for much of the early work that lead to the proposal of the ANNIE experiment. V. Fischer and S. Gardiner are supported by the DOE National Nuclear Security Administration through the Nuclear Science and Security Consortium under award number DE-NA0003180.

\bibliographystyle{model1-num-names}
\bibliography{bibliography.bib}

\end{document}